\documentclass[final,3p,times,authoryear]{elsarticle}

\usepackage{graphicx}          
\usepackage[dvips]{epsfig}    

\usepackage{ulem}
\usepackage{color}
\usepackage{braket}
\usepackage{amsmath,mathrsfs,amssymb}
\allowdisplaybreaks[4]
\usepackage{subeqnarray}
\usepackage{cases}
\usepackage{ntheorem}
\usepackage{amssymb}

\usepackage{wrapfig}
\usepackage{flushend}

\newtheorem{definition}{Definition}
\newtheorem{Theorem}{Theorem}
\newtheorem{corollary}{Corollary}
\newtheorem*{Proof}{Proof}

\journal{Automatica}

\begin{document}
\begin{frontmatter}
\title{Quantum feedback control of a two-atom network closed by a semi-infinite waveguide}

\author[Haijin Ding]{Haijin Ding} \ead{dhj17@tsinghua.org.cn}

\affiliation[Haijin Ding]{o={Department of Applied Mathematics}, a={The Hong Kong Polytechnic University}, c={Hung Hom}, p={Kowloon}, cy={Hong Kong}}

\author[Haijin Ding,Guofeng Zhang]{Guofeng Zhang\corref{cor1}}\ead{guofeng.zhang@polyu.edu.hk}

\affiliation[Guofeng Zhang]{o={The Hong Kong Polytechnic University Shenzhen Research Institute}, c = {Shenzhen}, p={518057}, cy={China}}

\author[Mu-Tian Cheng]{Mu-Tian Cheng} \ead{mtcheng@ahut.edu.cn}
\author[Mu-Tian Cheng]{Guoqing Cai}\ead{cgq268052@163.com}

\affiliation[Mu-Tian Cheng]{o={School of Electrical and Information Engineering}, a={Anhui University of Technology}, c={Maanshan}, p={243003}, cy={China}}

\cortext[cor1]{Corresponding author}


\begin{abstract}
The purpose of this paper is to study the delay-dependent coherent feedback dynamics by focusing on one typical realization, i.e., a two-atom quantum network whose feedback loop is closed by a semi-infinite waveguide. In this set-up, an initially excited two-level atom can emit one photon into the waveguide, where the propagating photon can be reflected by the terminal mirror of the waveguide or absorbed by the other atom, thus constructing  various coherent feedback loops. We show that there can be two-photon, one-photon or zero-photon states in the waveguide, which can be controlled by the feedback loop length and the coupling strengths between the atoms and waveguide. The photonic states in the waveguide  are analyzed in both the frequency domain and the spatial domain, and the transient process of photon emissions is better understood based on a comprehensive analysis using both domains. Interestingly, we clarify that this quantum coherent feedback network can be mathematically modeled as a linear control system with multiple delays, which are determined by the distances between atoms and the terminal mirror of the semi-infinite waveguide. Based on time-delayed linear control theory, the influence of delays on the stability of the quantum state evolution and the steady atomic or photonic states is investigated.
\end{abstract}

\end{frontmatter}


\section{Introduction}
\begin{sloppypar}
Quantum feedback control has found a variety of  applications in quantum information processing (QIP) and quantum engineering~\citep{Zhangjing2017}. According to whether the quantum state is measured, quantum feedback control can be divided into two categories:  measurement feedback control where the feedback control law is designed based on  the measurement results of the quantum state~\citep{Pierre2020exponential,YamamotoFeedbackDelay,YamamotoPRX} and coherent feedback control realized by coherent interactions among various quantum components in a quantum network ~\citep{zhang2010direct,zhang2020single,ZD22,entanglecavity}. Compared with measurement feedback control, one of the advantages of quantum coherent feedback control is that quantum states are not influenced by the measurement noises.  
\end{sloppypar}

\begin{sloppypar}
Among various quantum coherent feedback realizations, a most efficient approach is to construct coherent feedback channels using  waveguides. In waveguide quantum electro-dynamical (waveguide QED) systems, different components such as two-level systems~\citep{zhang2020dynamics,ZhangBin} or cavities~\citep{photonfeedback,crowder2020quantum} can be coupled to a waveguide, and photons transmitted in the waveguide can realize long-range interactions among the quantum nodes~\citep{simon2017towards,northup2014quantum,monroe2002quantum,flamini2018photonic}. Such quantum networks have been experimentally realized in platforms such as neutral atoms~\citep{neutralatom,threeLatom}, superconducting circuits~\citep{fly,peng2016tuneable}, trapped ions~\citep{keller2004continuous,ion,IonPRL}, and quantum dots~\citep{michler2000quantum,dotsingle}.
\end{sloppypar}

\begin{sloppypar}
Similar to classical multi-agent networks with time delays for control and communication among different agents~\citep{tao2022design,la2007asymptotic,li2016stabilization}, quantum coherent feedback control based on  photons propagating in a waveguide can be regarded as a networked control system with single~\citep{photonfeedback,ding2022quantum} or multiple time delays~\citep{ZhangBin,huo2020absorption,guimond2017delayed,ZollerPRL}. Mechanisms for the occurrence of time delays in quantum coherent feedback networks can be different for varied architecture designs. For example, in a  waveguide QED system where atoms or cavities are coupled to an infinite waveguide~\citep{zhang2020dynamics,WaveguideNomirror,chiralRoute,dinc2020diagrammatic,cheng2017waveguide,sinha2020collective,WaveguideModelingCan}, the feedback dynamics is only influenced by the transmission delays of photons among atoms or cavities. On the other hand, when the quantum nodes (e.g., atoms or cavities) are coupled to a semi-infinite waveguide~\citep{ZhangBin,ZollerPRL,WaveguideModelingCan}, the propagating photons can re-interact with the quantum nodes after being reflected by the terminal mirror of the waveguide, rendering an additional feedback channel influenced by the distance between the quantum nodes and the terminal mirror of the waveguide. From the perspective of control theory, traditional linear networked control systems with similar mathematical formats as the proposal above have attracted much attention, due to time delay's influence on the stability~\citep{xu2005improved,sun2010improved,shao2009new}, convergence rate~\citep{moradian2019positive}, robustness~\citep{kharitonov2002robust} and so on. Thus the dynamics of above quantum coherent feedback networks based on waveguide QED can be analyzed from the viewpoint of linear system with time delays.
\end{sloppypar}

\begin{sloppypar}
However, different from classical feedback control and quantum coherent feedback control with cavity quantum electro-dynamical (cavity QED) systems~\citep{lang1973laser}, a waveguide with continuous modes can provide a feedback channel with much larger spatial distribution compared with the size of quantum nodes, inducing interesting non-Markovian dynamics with time delays~\citep{photonfeedback,crowder2020quantum}, and this can be equivalently modeled in the spatial domain~\citep{Shentheory1,boundary,PRLDuke,yan2011controlling,zhou2017single,InfiniteWaveguidePFeq}.  
Then the photons propagate in the format of wave packets can be characterized by their spatial distributions in the waveguide and the wave packets of single-photon and multi-photon states evolve according to waveguide's coupling to other quantum nodes~\citep{ShentwophtonP,bradford2015architecture,cheng2017waveguide}.
Above all, a comprehensive analysis in the frequency and spatial domains should be a most desirable approach to investigate the coherent feedback control dynamics in waveguide QED.
\end{sloppypar}

\begin{sloppypar}
Depending on whether the coupling strengths between a quantum node (i.e., an atom or a cavity) and directional photonic fields in the waveguide are identical or not, the interaction between the quantum node and the waveguide can be categorized to be nonchiral or chiral~\citep{lodahl2017chiral}. One representative example of such systems is the architecture in Fig.~\ref{fig:scheme}, where two two-level atoms are coupled to a semi-infinite waveguide with at most two excitations. In this set-up, the evolution of the quantum states is simultaneously influenced by the coherent feedback interactions between two atoms as in ~\citep{zhang2020dynamics}, the coherent feedback induced by the mirror as in ~\citep{waveguideOneatom,boundary}, and the chiral couplings between two atoms and a semi-infinite waveguide as in ~\citep{ZhangBin}. By analyzing the dynamics of this coherent feedback network in both the frequency and the spatial domains from the perspective of linear control systems with multiple delays, we show that the steady atomic and photonic states can be different depending on coherent feedback designs.
\end{sloppypar}

\begin{sloppypar}
The rest of the paper is organized as follows. Section~\ref{Sec:Model} analyzes in the frequency domain on the coherent feedback dynamics where a semi-infinite waveguide is coupled to two initially excited two-level atoms, especially on how the chiral couplings and time delays induced by atoms' positions can influence the performance of the coherent feedback network. Section~\ref{Sec:position} presents the analysis in the spatial domain and demonstrates how a single-photon wave packet propagates in the two-atom network mediated by a waveguide. Section~\ref{Sec:conclusion} concludes this paper.
\end{sloppypar}

\emph{Notation.} The reduced Planck constant $\hbar$,  the velocity of the light field $c$ and the group velocity of the propagating photonic wave packets $v_g$ are set to $1$ in this paper.  

\section{Coherent feedback dynamics of two atoms coupled to a semi-infinite waveguide}   \label{Sec:Model}
\begin{figure}[h]
\centerline{\includegraphics[width=0.8\columnwidth]{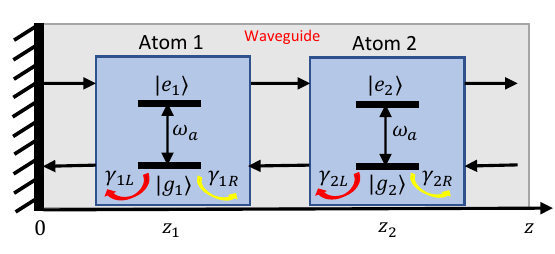}}
\caption{A quantum coherent feedback control network where two atoms are coupled to a semi-infinite waveguide.}
	\label{fig:scheme}
\end{figure}
As shown in Fig.\ref{fig:scheme}, two atoms (or atomlike objects) with the identical resonant frequency $\omega_a$ between two energy levels, i.e., $|g_j\rangle$ and $|e_j\rangle$ with $j=1,2$, are coupled to a semi-infinite waveguide. The real-value coupling strength between the $j$th atom and the left-propagating field in the waveguide is $\gamma_{jL}$ , and that for the right propagating field is $\gamma_{jR}$. For the atom at $z_1$, the emitted field in the left-propagating direction can  be reflected by the mirror at $z=0$ and then propagates along the right direction to re-interact with the atom, whereas the emitted field in the right-propagating direction excites the atom at $z_2$.   For the atom at $z_2$, the emitted filed in the left-propagating direction interacts with the atom at $z_1$, while the emitted filed in the right-propagating direction leaves the system. The following assumption is adopted in this section.
\newtheorem{assumption}{Assumption}
\begin{assumption} \label{Initial}
The two atoms are both initially excited, the waveguide is empty,  and there are no external drives.  
\end{assumption}

The free Hamiltonian of the system reads
\begin{equation} \label{con:H0}
\begin{aligned}
H_0 &=  \sum_{j=1,2} \omega_a \sigma_j^+ \sigma_j^- + \int    \omega d^{\dag}_kd_k \mathrm{d}k ,
\end{aligned}
\end{equation}
where the first component represents the atomic Hamiltonian, and the second component represents the waveguide Hamiltonian. Here, $\sigma_j^- = |g_j\rangle \langle e_j |$ and $\sigma_j^+= |e_j\rangle \langle g_j|$ are the lowering and raising operators of the $j$th atom respectively, $d_{k}$($d_{k}^{\dag}$) are the annihilation(creation) operators of the propagating waveguide modes, $\omega = ck$ and $c$ is the velocity of the field in the waveguide.

\begin{sloppypar}
The interaction Hamiltonian between the atoms and waveguide in the interaction picture is~\citep{WaveguideModelingCan,AtomWaveguideHam,halfcavity,ZollerPRL,ZhangBin}
\end{sloppypar}
\begin{equation} \label{con:HintjOrig}
\begin{aligned}
H_{\rm I} = & i \sum_{j=1,2}\int  \left[\left (d^{\dag}_{k}\sigma_j^-\gamma_{jR}e^{-i\omega z_j/c}e^{-i\Phi/2}e^{i(\omega-\omega_a)t} \right.\right.\\
&\left.\left.- d^{\dag}_{k}\sigma_j^-\gamma_{jL}e^{i\omega z_j/c}e^{i\Phi/2}e^{i(\omega-\omega_a)t}\right) - \mathrm{H.c.}\right]\mathrm{d}k,
\end{aligned}
\end{equation}
where $\Phi$ is the phase induced by  the mirror reflection which is experimentally small~\citep{boundary}, $d^{\dag}_{k}\sigma_j^-$ represents that an excited atom at $z_j$ can emit photon with the mode $k$ via decaying to its ground state, and $\mathrm{H.c.}$ denotes Hermitian conjugate representing the reverse process. The continuous coupling modes between the atoms and the waveguide are integrated within $[0,+\infty)$ in this paper. 

Considering that the phase shift $\Phi$ vanishes under the global translation of the atoms' positions, the interaction Hamiltonian in the interaction picture can be simplified as
\begin{equation} \label{con:Hintj}
H_{\rm I} = \sum_{j=1,2}\int  \left[g_{kjt}(k,t,z) d^{\dag}_k\sigma_j^-  + g_{kjt}^*(k,t,z) d_k\sigma_j^+\right]\mathrm{d}k , 
\end{equation}
where
\begin{equation}
\begin{aligned} 
g_{kjt}(k,t,z)&\triangleq i\left(\gamma_{jR} e^{-i\omega z_j/c}-\gamma_{jL} e^{i\omega z_j/c}\right) e^{i\left(\omega-\omega_a\right)t}.
\label{eq:g_kjt}
\end{aligned}
\end{equation}
When the couplings between the atoms and the waveguide are {\it nonchiral}, namely $\gamma_{jR} = \gamma_{jL} \equiv \gamma_{j}$ with $j=1,2$, the coupling strength in Eq. \eqref{eq:g_kjt} reduces to $g_{kjt} = 2\gamma_j \sin(kz_j) e^{i(\omega-\omega_a)t}$.

According to \textbf{Assumption \ref{Initial}}, the quantum coherent feedback network has at most two excitations, hence the state vector of the system can be represented as
\begin{small}
\begin{equation} \label{con:State}
\begin{aligned}
&|\Psi(t)\rangle = c_{ee}(t) |e_1,e_2,\{0\}\rangle + \int c_{egk}(t,k) |e_1,g_2,\{k\}\rangle \mathrm{d}k \\
&+ \int c_{gek}(t,k) |g_1,e_2,\{k\}\rangle \mathrm{d}k +  \iint c_{kk}(k_1,k_2,t) |g_1,g_2,\{k_1\}\{k_2\}\rangle  \mathrm{d}k_1\mathrm{d}k_2.
\end{aligned}
\end{equation}
\end{small}%
Here, $|e_1,e_2,\{0\}\rangle$ represents that both atoms are excited and there are no photons in the waveguide, $|e_1,g_2,\{k\}\rangle$ indicates that the first atom is excited, the second atom is in its ground state, and there is one photon with the mode $k$ in the waveguide, $|g_1,e_2,\{k\}\rangle$ represents that the first atom is in its ground state, the second atom is excited and there is one photon in the waveguide, and finally, $|g_1,g_2,\{k_1\}\{k_2\}\rangle$ represents that both atoms are in their ground states and there are two photons with modes $k_1$ and $k_2$ in the waveguide.  $c_{ee}(t)$, $c_{egk}(t,k)$, $c_{gek}(t,k)$ and  $c_{kk}(k_1,k_2,t)$ represent the quantum state amplitudes respectively.

\begin{sloppypar}
The quantum state vector dynamics is governed by the Schr\"{o}dinger equation in the interaction picture as
\begin{equation} \label{con:Schrequation}
\begin{aligned}
\frac{\mathrm{d}}{\mathrm{d} t}|\Psi(t)\rangle = -i H_{\rm I} |\Psi(t)\rangle,
\end{aligned}
\end{equation}
with the Hamiltonian in Eq. \eqref{con:Hintj} and  the ansatz in Eq. \eqref{con:State}, yields a system of integro-differential equations for the amplitudes
\begin{subequations} \label{con:Popuquation}
\begin{align}
&\dot{c}_{ee}(t) = - i\int c_{egk}(t,k)g_{k2t}^*(k,t,z_2)\mathrm{d}k \notag\\
&~~~~~~~~~~~~ -i \int c_{gek}(t,k) g_{k1t}^*(k,t,z_1)\mathrm{d}k , \label{model1}\\
&\dot{c}_{egk}(t,k) = -i c_{ee}(t)  g_{k2t}(k,t,z_2)  \notag\\
&~~~~~~~~~~~~~~~~~ - i \int c_{kk}(k,k_1,t)g_{k1t}^*(k_1,t,z_1) \mathrm{d}k_1 ,\label{model2}\\
&\dot{c}_{gek}(t,k) = -i c_{ee}(t) g_{k1t}(k,t,z_1) \notag\\
&~~~~~~~~~~~~~~~~~-i \int c_{kk}(k,k_1,t)g_{k2t}^*(k_1,t,z_2) \mathrm{d}k_1 ,\label{model3}\\
&\dot{c}_{kk}(k_1,k_2,t) = -i c_{egk}(t,k_1)  g_{k1t}(k_2,t,z_1)  -i c_{egk}(t,k_2)  g_{k1t}(k_1,t,z_1) \notag\\
&- i c_{gek}(t,k_1) g_{k2t}(k_2,t,z_2)   - i c_{gek}(t,k_2) g_{k2t}(k_1,t,z_2). \label{model4}
\end{align}
\end{subequations}
The physical interpretation of Eq. \eqref{con:Popuquation} is as follows.  Eq.~(\ref{model1}) means that if one of the two atoms is excited while the other is in its ground state, both of the two atoms can be excited by absorbing one photon from the waveguide. Conversely, the first component at the right-hand side of Eq.~(\ref{model2}) or Eq.~(\ref{model3}) indicates that when the two atoms are both excited, one of the atoms can emit one photon into the waveguide and decay to its ground state, whereas the second item  describes that one atom can absorb one photon from the waveguide if both of the two atoms are in their ground states. Eq.~(\ref{model4}) represents that when only one of the two atoms is excited, the excited atom can emit one photon into the waveguide, then both atoms are in there ground states and there are two photons in the waveguide.
\end{sloppypar}

According to the methods introduced in \textbf{\ref{Sec:ChiralOdecal}},  Eq. (\ref{model1}) can be rewritten as
\begin{equation} \label{con:Ceedelay}
\begin{aligned}
&\dot{c}_{ee}(t) =-\frac{\gamma_{1R}^2 + \gamma_{1L}^2 + \gamma_{2R}^2 + \gamma_{2L}^2 }{2}  c_{ee}(t)\\
&+ \gamma_{1L}\gamma_{1R} c_{ee}\left(t-\frac{2z_1}{c}\right)e^{i\omega_a\frac{2z_1}{c}}  + \gamma_{2L}\gamma_{2R} c_{ee}\left(t-\frac{2z_2}{c}\right) e^{i\omega_a\frac{2z_2}{c}},
\end{aligned}
\end{equation}
which demonstrates the influence by the round-trip time delays between the atoms and the terminal mirror. Similarly, Eq.~(\ref{model2}) can be rewritten as
\begin{equation} \label{con:Cegdelay}
\begin{aligned}
\dot{c}_{egk}(t,k) =&-\frac{\gamma_{1R}^2  + \gamma_{1L}^2 }{2} c_{egk}(t,k) -i c_{ee}(t)  g_{k2t}(k,t,z_2) \\
&+ \gamma_{1L} \gamma_{1R} c_{egk}\left(t-\frac{2z_1}{c},k\right) e^{i\omega_a \frac{2z_1}{c}}\\
& +\gamma_{1R}\gamma_{2L} c_{gek}\left(t-\frac{z_1+z_2}{c},k\right) e^{i\omega_a \frac{z_1+z_2}{c}}\\
& - \gamma_{1L}\gamma_{2L} c_{gek}\left(t-\frac{z_2-z_1}{c},k\right) e^{i\omega_a \frac{z_2-z_1}{c}}, 
\end{aligned}
\end{equation}
where $\left(z_1+z_2\right)/c$ represents the time delay from the second atom at $z_2$ to the first atom at $z_1$ after the coherent field is reflected by the mirror, and $\left(z_2-z_1\right)/c$ represents the time delay directly from the second atom at $z_2$ to the first atom at $z_1$. Finally, Eq.~(\ref{model3}) can be rewritten as
\begin{equation} \label{con:Cgedelay}
\begin{aligned}
\dot{c}_{gek}(t,k) =&- \frac{\gamma_{2R}^2 + \gamma_{2L}^2}{2}  c_{gek}(t,k)  -i c_{ee}(t) g_{k1t}(k,t,z_1)  \\
&+ \gamma_{2L}\gamma_{2R}  c_{gek}\left(t-\frac{2z_2}{c},k\right) e^{i\omega_a \frac{2z_2}{c}}  \\
 & + \gamma_{1L} \gamma_{2R}   c_{egk}\left(t-\frac{z_1+z_2 }{c},k\right)  e^{i\omega_a \frac{z_1+z_2}{c}} \\
 &- \gamma_{1R}\gamma_{2R}   c_{egk}\left(t-\frac{z_2-z_1}{c},k\right) e^{i\omega_a\frac{z_2-z_1}{c}},
\end{aligned}
\end{equation}
where $\left(z_1+z_2\right)/c$ and $\left(z_2-z_1\right)/c$ represent the time delay from the first atom at $z_1$ to the second atom at $z_2$ via the path reflected by the mirror or direct transmission, respectively.

In particular, if we take $\gamma_{jL} = \gamma_{jR}$ for $j=1,2$,  the above equations reduce to the nonchiral coupling circumstance. For simplicity, we denote
\begin{equation} \label{con:gLRSimplify}
\gamma_{RL} = \frac{\gamma_{1R}^2 + \gamma_{1L}^2 + \gamma_{2R}^2 + \gamma_{2L}^2 }{2},  \  \mathbf{g}_j = \gamma_{jL}\gamma_{jR}, \  \tau_j = \frac{2z_j}{c},
\end{equation}
where $j =1,2$. Obviously,
$\gamma_{RL}  \geq  \mathbf{g}_1+ \mathbf{g}_2$,
and the equality holds only when $\gamma_{1L} = \gamma_{1R}$ and $\gamma_{2L} = \gamma_{2R}$. Applying the Laplace transformation to both sides of  Eq.~(\ref{con:Ceedelay}), 
we get
\begin{equation} \label{eq:sep23_C_ee}
\begin{aligned}
 &C_{ee}(s) =\frac{c_{ee}(0)}{s+ \gamma_{RL} -\mathbf{g}_1 e^{i\omega_a\tau_1} e^{-\tau_1s} -  \mathbf{g}_2 e^{i\omega_a\tau_2} e^{-\tau_2 s}}.
 \end{aligned}
\end{equation}

Next, we investigate the dynamics of the coherent feedback network based on the following assumption.

\begin{assumption} \label{Markovian}
The resonant frequency of the atoms satisfies $\omega_a \gg 1$, and atoms' positions satisfy that $z_j\ll c$ for $j =1,2$.
\end{assumption}
  
The inverse Laplace transformation of  Eq.~(\ref{eq:sep23_C_ee}) is taken by integrating on the positive half of the complex plane close to the imaginary axis. By \textbf{Assumption~\ref{Markovian}}, $e^{-2z_j s/c}\approx 1 $ and the amplitude $c_{ee}(t)$  can be approximated as
\begin{equation} \label{con:Ceetapprox}
\begin{aligned}
c_{ee}(t)  & \approx e^{\left[-\gamma_{RL} +  \gamma_{1L}\gamma_{1R} \cos\left(\omega_a\frac{2z_1}{c}\right) + \gamma_{2L}\gamma_{2R}\cos\left(\omega_a\frac{2z_2}{c}\right)\right]t}\\
&~~~~e^{i\gamma_{1L}\gamma_{1R} \sin\left(\omega_a\frac{2z_1}{c}\right)t}
e^{i\gamma_{2L}\gamma_{2R}\sin\left(\omega_a\frac{2z_2}{c}\right)t}.
\end{aligned}
\end{equation}

Because $\omega_a \gg 1$,  the value $\omega_a z_j/c$ with $j=1,2$ varies in the interval $[0, n\pi]$ for some positive integer $n$ even when $z_j\ll c$. Consequently, the quantum coherent feedback network dynamics can be distinctly influenced by $z_j$, which is studied in detail in the following subsections.

\subsection{Control the photon numbers in the waveguide by means of time delays and chiral couplings}
\begin{sloppypar}
Based on the delay-dependent control equations  (\ref{con:Ceedelay},\ref{con:Cegdelay},\ref{con:Cgedelay}), the number of photons in the waveguide can be zero, one or two according to the parameter settings, due to \textbf{Theorems \ref{chiralsteady}}-\textbf{\ref{Promatain}} in the following.
\end{sloppypar}

\begin{Theorem} \label{chiralsteady}
Under \textbf{Assumption~\ref{Markovian}},
$\lim_{t\rightarrow \infty} c_{ee}(t)= 0$ when the coupling between the waveguide and at least one of the two atoms is chiral.
\end{Theorem}
\begin{Proof}
When $\gamma_{jL} \neq \gamma_{jR}$ for $j=1$ or 2, 
\begin{equation} \label{con:Inequal}
\begin{aligned}
\gamma_{RL} &> \gamma_{1L}\gamma_{1R} + \gamma_{2L}\gamma_{2R} \\
&\geq \gamma_{1L}\gamma_{1R}\cos\left(\omega_a\frac{2z_1}{c}\right)+ \gamma_{2L}\gamma_{2R}\cos\left(\omega_a\frac{2z_2}{c}\right).
\end{aligned}
\end{equation}
Thus for arbitrary $\gamma_{jL}$, $\gamma_{jR}$ and $z_j$,
\begin{small}
\begin{equation} \label{con:Ceeproof}
\begin{aligned}
-\gamma_{RL} +  \gamma_{1L}\gamma_{1R} \cos\left(\omega_a\frac{2z_1}{c}\right) + \gamma_{2L}\gamma_{2R}\cos\left(\omega_a\frac{2z_2}{c}\right) < 0.
\end{aligned}
\end{equation}
\end{small}
Then by Eq. \eqref{con:Ceetapprox}, $\lim_{t\rightarrow \infty} c_{ee}(t)= 0$.
\qed
\end{Proof}

\begin{figure}[h]
\centerline{\includegraphics[width=1\columnwidth]{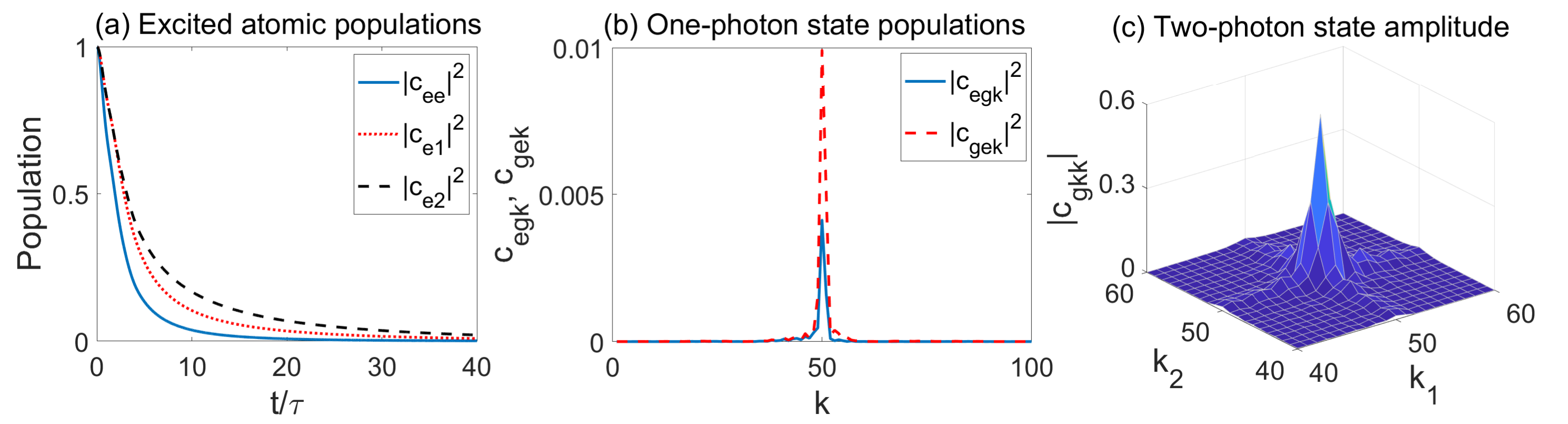}}
\caption{The evolution of two atoms coupled to a waveguide evaluated by the populations of excited atomic states in (a), the populations of quantum states with one photon in the waveguide in (b), and the amplitude of quantum state with two photons in the waveguide in (c).}
	\label{fig:nochiral}
\end{figure}

Let $|c_{ej}(t)|^2$ with $j =1,2$ denote the population that the $j$th atom is excited, namely
\begin{equation} \label{con:Ce1Ce2}
\left\{
\begin{aligned}
&|c_{e1}(t)|^2 = |c_{ee}(t)|^2+ \int |c_{egk}(t,k)|^2 \mathrm{d}k, \\
&|c_{e2}(t)|^2 = |c_{ee}(t)|^2+ \int |c_{gek}(t,k)|^2 \mathrm{d}k.\\
\end{aligned}
\right.
\end{equation}
In the numerical simulations in Fig.~\ref{fig:nochiral}, $\gamma_{jR}=2\gamma_{jL} = 0.5$, $\omega_a = 50$, $z_1 = 0.1$, $z_2 = 0.2$, and $\tau= z_1/c$. Fig.~\ref{fig:nochiral}(a) shows that the populations of the excited states finally converge to zero. Moreover, the populations of the single-photon states also converge to zero, as shown in Fig.~\ref{fig:nochiral}(b) when $t=40\tau$. As a result, there are two photons in the waveguide and the population of the two-photon state with the modes $k_1$ and $k_2$ is shown in Fig.~\ref{fig:nochiral}(c) when $t=40\tau$.

For typical chiral interactions between the atoms and the waveguide,  by \textbf{Theorem \ref{chiralsteady}} asymptotically there are two photons in the waveguide and the atoms are in their ground states. However, under certain extreme conditions,  it may happen that there is only one photon in the waveguide and one atom is in the excited state persistently, as given in the following theorem.

\begin{Theorem} \label{onephoton} When $\gamma_{1R} = \gamma_{1L}$, $z_1= n\pi/\omega_a \ll c$ for some positive integer $n$,  $z_2 \ll c$, $\gamma_{2R} > \gamma_{2L}= 0$ or $\gamma_{2L} > \gamma_{2R}= 0$, the first atom  holds a significant
amount of excitation and the second atom decays to the ground state. 
\end{Theorem}

\begin{Proof}
We look at the case $\gamma_{2L} > \gamma_{2R} = 0$. The other case $\gamma_{2L} > \gamma_{2R}= 0$ can be proved similarly.  Under the assumptions in \textbf{Theorem \ref{onephoton}},  Eqs.~(\ref{con:Ceedelay},\ref{con:Cegdelay},\ref{con:Cgedelay})  become
\begin{subequations}  \label{con:NonchiralControlChiralG2R0}
\begin{align}
&\dot{c}_{ee}(t) =-\frac{\gamma_{2L}^2 + \gamma_{1R}^2 + \gamma_{1L}^2}{2}  c_{ee}(t) + \gamma_{1L}\gamma_{1R} c_{ee}\left(t-\frac{2z_1}{c}\right)e^{i\omega_a\frac{2z_1}{c}}, \\
&\dot{c}_{gek}(t,k) = -i c_{ee}(t) g_{k1t}(k,t,z_1)    - \frac{\gamma_{2L}^2}{2}  c_{gek}(t,k), \\
&\dot{c}_{egk}(t,k) =  -i c_{ee}(t)  g_{k2t}(k,t,z_2) -\frac{\gamma_{1R}^2  + \gamma_{1L}^2 }{2} c_{egk}(t,k) \notag\\
&~~~~~~~~~~~~~~~~~+ \gamma_{1L} \gamma_{1R} c_{egk}\left(t-\frac{2z_1}{c},k\right) e^{i\omega_a \frac{2z_1}{c}}\notag\\
&~~~~~~~~~~~~~~~~~+\gamma_{1R}\gamma_{2L} c_{gek}\left(t-\frac{z_1+z_2}{c},k\right) e^{i\omega_a \frac{z_1+z_2}{c}}\notag\\
&~~~~~~~~~~~~~~~~~- \gamma_{1L}\gamma_{2L} c_{gek}\left(t-\frac{z_2-z_1}{c},k\right) e^{i\omega_a \frac{z_2-z_1}{c}}.
\end{align}
\end{subequations}

Applying the Laplace transformation to the first two equations of (\ref{con:NonchiralControlChiralG2R0}) yields
\begin{equation} \label{con:LaplaceCee}
\begin{aligned}
C_{ee}(s)= \frac{1}{s+ \frac{\gamma_{2L}^2 + \gamma_{1R}^2 + \gamma_{1L}^2}{2} - \gamma_{1L}\gamma_{1R}  e^{i\omega_a\frac{2z_1}{c}} e^{-\frac{2z_1}{c}s}}.
\end{aligned}
\end{equation}
By the final-value theorem,
\begin{equation} \label{con:Ceetinfinite}
\begin{aligned}
&~~~~\lim_{t\rightarrow \infty}c_{ee}(t)=\lim_{s\rightarrow 0}sC_{ee}(s)\\
&= \lim_{s\rightarrow 0}\frac{s}{s+ \frac{\gamma_{2L}^2 + \gamma_{1R}^2 + \gamma_{1L}^2}{2} - \gamma_{1L}\gamma_{1R}  e^{i\omega_a\frac{2z_1}{c}} e^{-\frac{2z_1}{c}s}} =0 .
\end{aligned}
\end{equation}

\begin{sloppypar}
As a result, in the long-time limit, $\dot{c}_{gek}(t,k) \approx - \left(\gamma_{2L}^2/2\right)  c_{gek}(t,k)$, and hence $\lim_{t\rightarrow\infty}c_{gek}(t,k) = 0$. Thus, if $t>t_1$ for some $t_1$ large enough,  $c_{ee}(t) \approx 0$ and $c_{gek}(t,k) \approx 0$.   In this case, when $z_1 = n\pi/\omega_a$, the third equation of Eq.~(\ref{con:NonchiralControlChiralG2R0}) can be simplified when $t>t_1$ as
\end{sloppypar}
\begin{small}
\begin{equation}\label{con:cegksim}
\begin{aligned}
\dot{c}_{egk}(t,k) \approx & -\frac{\gamma_{1R}^2  + \gamma_{1L}^2 }{2} c_{egk}(t,k) + \gamma_{1L} \gamma_{1R} c_{egk}\left(t-\frac{2z_1}{c},k\right) e^{i\omega_a \frac{2z_1}{c}}.
\end{aligned}
\end{equation}
\end{small}%
The Laplace transformation of Eq.~(\ref{con:cegksim}) yields
\begin{equation}
\begin{aligned}
C_{egk}(s,k) \approx \frac{ c_{egk}(t_1,k)}{s+\frac{\gamma_{1R}^2  + \gamma_{1L}^2 }{2} - \gamma_{1L} \gamma_{1R}e^{i\omega_a \frac{2z_1}{c}}}.
\end{aligned}
\end{equation}
Consequently, when $\gamma_{1L} = \gamma_{1R}$ and $\omega_a z_1/c = n\pi$, $c_{egk}(t,k) \approx  c_{egk}(t_1,k)$.
\qed
\end{Proof}

\begin{figure}[h]
\centerline{\includegraphics[width=1\columnwidth]{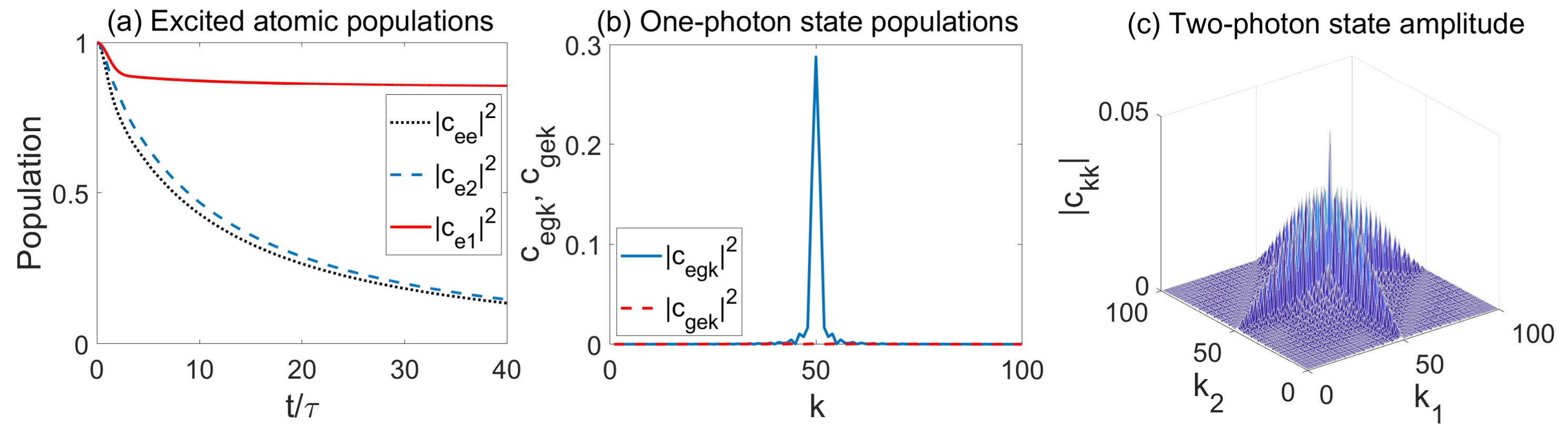}}
\caption{Control performance for the generation of single-photon states evaluated by the populations of excited atomic states in (a), the populations of quantum states with one photon in the waveguide in (b), and the amplitude two-photon states in the waveguide in (c).}
	\label{fig:onephotons}
\end{figure}

The numerical simulations in Fig. \ref{fig:onephotons} agree with \textbf{Theorem}~\ref{onephoton}, where $z_1 = \pi/\omega_a$, $z_2 = 2\pi/\omega_a$, $\gamma_{1L} = \gamma_{1R} = 0.25$, $\gamma_{2L} = 0.5$, $\gamma_{2R} = 0$, and $\omega_a = 50$. As shown in Fig. \ref{fig:onephotons}(a), the population $|c_{ee}(t)|^2$   converges to zero, while  the population $|c_{e1}(t)|^2$ that the first atom is excited  remains around $0.86$. As in Fig. \ref{fig:onephotons}(b), the amplitude of the single-photon state $c_{egk}$ approximates a bump function at $\omega=\omega_a$ and $t=40\tau$. Finally, as shown in Fig.~\ref{fig:onephotons}(c), there are no obvious two-photon states in the waveguide when $t=40\tau$.

\newtheorem{remark}{Remark}

For the nonchiral case, we have the following result.

\begin{Theorem} \label{Promatain}
For the nonchiral coupling between two atoms and the semi-infinite waveguide, $c_{ee}(t) \approx 1$ when $z_j = n\pi /\omega_a \ll c$ with $n=0,1,2,\cdots$ with $j=1,2$.
\end{Theorem}
\begin{Proof}
\begin{sloppypar}
For the nonchiral coupling circumstance with $\gamma_{jR} = \gamma_{jL}$, when $z_j = n\pi /\omega_a \ll c$ with $n=0,1,2,\cdots$, $-\gamma_{RL} +  \gamma_{1L}\gamma_{1R} \cos\left(2\omega_a z_1/c\right) + \gamma_{2L}\gamma_{2R}\cos\left(2\omega_a z_2/c\right)\approx 0$ due to $\sin\left(2\omega_a z_j/c\right) = 0$ and $\cos\left(2\omega_a z_j/c\right) = 1$ in Eq.~(\ref{con:Ceetapprox}). As a result, $c_{ee}(t) \approx 1$.
\qed
\end{sloppypar}
\end{Proof}

\begin{figure}[h]
\centerline{\includegraphics[width=0.7\columnwidth]{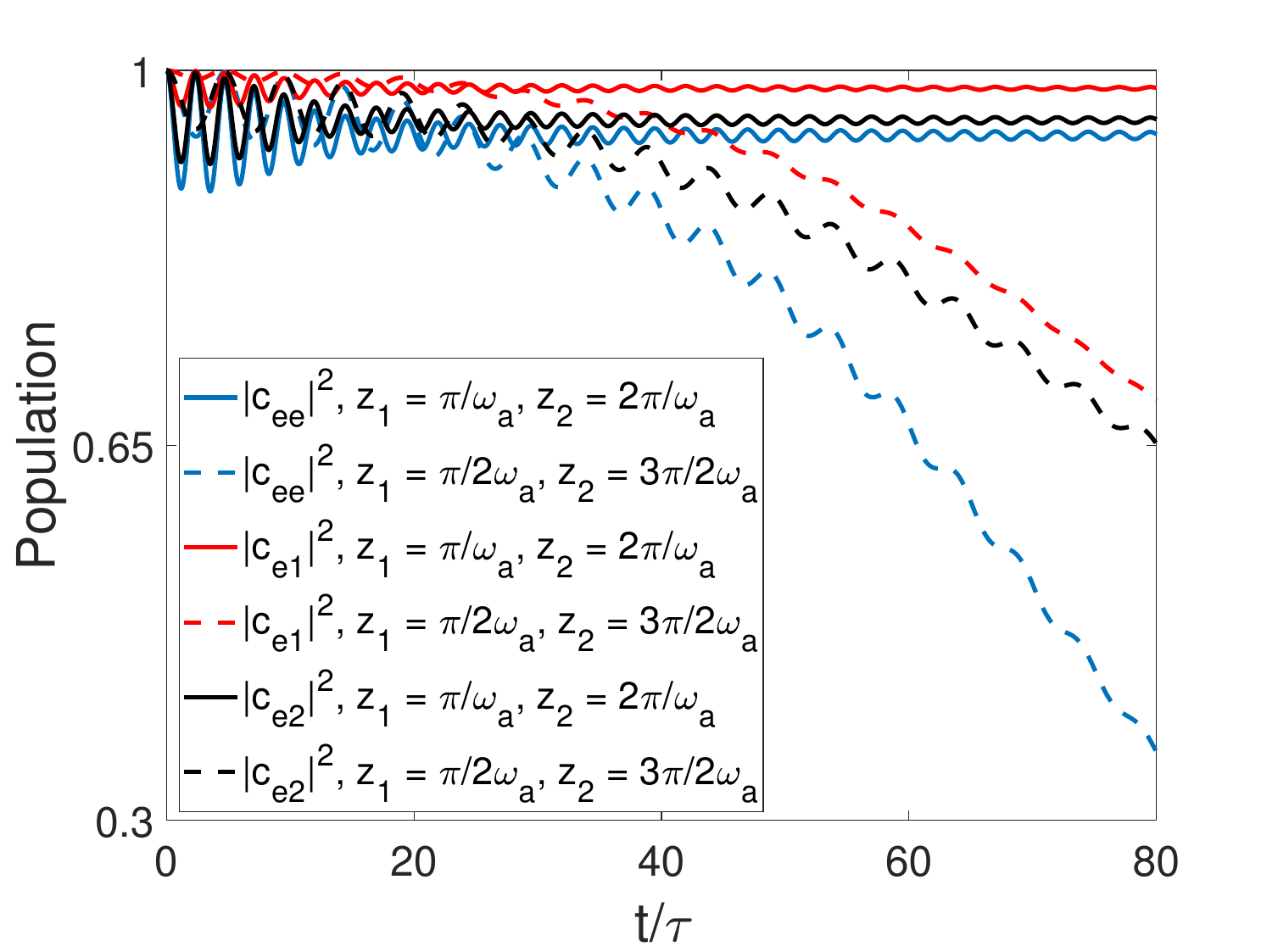}}
\caption{Comparison of the excited populations of two atoms influenced by atoms' positions.}
	\label{fig:locationcontrol}
\end{figure}

As compared in Fig.\ref{fig:locationcontrol}, we take $\omega_a = 50$, $\gamma_{jR} = \gamma_{jL} = 0.5$ with  $j=1,2$ for both of the two atoms. The comparisons agree with the conclusions in  
\textbf{Theorem~\ref{chiralsteady}} and \textbf{Theorem~\ref{Promatain}}. That is, when the positions of the two charily coupled atoms are $z_j = n\pi /\omega_a$ with $n=1,2,\cdots$, which is the case for the solid lines in Fig.\ref{fig:locationcontrol}, the two atoms maintain their excited states due to the coherent feedback interactions.

In summary, for the coherent feedback network where two atoms are coupled to a semi-infinite waveguide, the number of photons in the waveguide can be controlled by tuning the chiral coupling strengths between the atoms and waveguide as well as atoms' positions, which makes it possible to generate zero-photon, one-photon and two-photon states in the waveguide.

\subsection{Relationship between coherent feedback network dynamics and time delays}
In this subsection, we study the relationship between time delays and atomic dynamics from two perspectives; one is based on a quasi-polynomial model by applying the Laplace transformation to the linear control equation with time delays, and the other is based on the master equation within Markovian approximation, where the time delays play the role of phase modulations.

\subsubsection{Quasi-polynomial model} 
\begin{sloppypar}
Consider the time delay dependent Eq.~(\ref{con:Ceedelay}), which can be complex-valued for most parameter settings of $z_1$, $z_2$ and $\gamma_{jL}$, $\gamma_{jR}$ with $j=1,2$. We rewrite this equation by denoting the real and imaginary parts of $c_{ee}(t)$ as real-valued functions $c_{ee}^R(t)$ and $c_{ee}^I(t)$, respectively. Namely, $c_{ee}(t) = c_{ee}^R(t) + ic_{ee}^I(t)$. According to Eq.~(\ref{con:gLRSimplify}), we have
\begin{subequations} \label{con:CeeRealImag}
\begin{align}
&\dot{c}_{ee}^R(t) =-\gamma_{RL}c_{ee}^R(t) + \mathbf{g}_1c_{ee}^R\left(t-\tau_1\right)\cos\left(\omega_a\tau_1\right) \notag\\
&~~~~-\mathbf{g}_1 c_{ee}^I\left(t-\tau_1\right)\sin\left(\omega_a\tau_1\right)+ \mathbf{g}_2 c_{ee}^R\left(t-\tau_2\right)\cos\left(\omega_a\tau_2\right)\notag\\
&~~~~-\mathbf{g}_2 c_{ee}^I\left(t-\tau_2\right)\sin\left(\omega_a\tau_2\right), \label{ceeRdot}\\
&\dot{c}_{ee}^I(t) =-\gamma_{RL}c_{ee}^I(t) + \mathbf{g}_1 c_{ee}^R\left(t-\tau_1\right)\sin\left(\omega_a\tau_1\right) \notag\\
&~~~~+\mathbf{g}_1 c_{ee}^I\left(t-\tau_1\right)\cos\left(\omega_a\tau_1\right)+ \mathbf{g}_2 c_{ee}^R\left(t-\tau_2\right)\sin\left(\omega_a\tau_2\right)\notag\\
&~~~~+\mathbf{g}_2 c_{ee}^I\left(t-\tau_2\right)\cos\left(\omega_a\tau_2\right), \label{ceeIdot}
\end{align}
\end{subequations}
according to the definition of variables in Eq.~(\ref{con:gLRSimplify}). We treat the delay-dependent components 
\begin{equation} \label{con:udelay}
\left\{
\begin{aligned}
&u_1(t) =c_{ee}^R\left(t-\tau_1\right),~~v_1(t) =c_{ee}^I\left(t-\tau_1\right),\\
&u_2(t) =c_{ee}^R\left(t-\tau_2\right),~~v_2(t) =c_{ee}^I\left(t-\tau_2\right),
\end{aligned}
\right.
\end{equation}
as the controls with time delays. Denote $\mathbf{x}(t) = \left[c_{ee}^R(t),c_{ee}^I(t) \right]^{\rm T}$ with $\rm T$ representing the transpose of a matrix, $\mathbf{u}_1 = \left[ u_1,v_1\right]^{\rm T}$ and $\mathbf{u}_2 = \left[u_2,v_2\right]^{\rm T}$. Then Eq.~(\ref{con:CeeRealImag}) can be rewritten with a format a linear control system as
\begin{equation} \label{con:linearControlxt}
\dot{\mathbf{x}}(t) = A\mathbf{x}(t) + B_1\mathbf{u}_1(t)  + B_2\mathbf{u}_2(t),
\end{equation}
where $A = {\rm diag}\left(-\gamma_{RL}  ,-\gamma_{RL} \right) $,
\begin{equation}
\begin{aligned}
  B_j &= \begin{bmatrix}
    \mathbf{g}_j\cos\left(\omega_a\tau_j\right) & -\mathbf{g}_j\sin\left(\omega_a\tau_j\right)  \\
   \mathbf{g}_j\sin\left(\omega_a\tau_j\right) & \mathbf{g}_j\cos\left(\omega_a\tau_j\right)  \\
  \end{bmatrix}, \ \  j=1,2.
  \end{aligned}
\end{equation}
It can be readily shown that the characteristic equation of  the   system \eqref{con:linearControlxt} reads~\citep{angulo2019quasi}
\begin{equation} \label{con:linearControlxtLast}
\begin{aligned}
\Delta_{c} (s) &= \left | sI -A -B_1 e^{-\tau_1 s} -B_2 e^{-\tau_2 s} \right|\\
&=\left( s+ \gamma_{RL}- \sum_j \mathbf{b}_je^{-\tau_j s}\right)^2 +\left(\sum_j \tilde{\mathbf{b}}_j e^{-\tau_j s}  \right)^2,
\end{aligned}
\end{equation}
where  $\mathbf{b}_j = \mathbf{g}_j\cos\left(\omega_a\tau_j\right)= \gamma_{jL}\gamma_{jR}\cos\left(\omega_a\tau_j\right)$ and $\tilde{\mathbf{b}}_j = \mathbf{g}_j\sin\left(\omega_a\tau_j\right)$. 
Clearly, when $\tau_j \approx 0$ for $j=1,2$, Eq. \eqref{con:linearControlxtLast} can be approximated as
\begin{equation} \label{con:DeltasSCase1}
\begin{aligned}
&\Delta_{c} (s) 
\approx \left( s+ \gamma_{RL}- \sum_j \mathbf{b}_j\right)^2 +\left(\sum_j \tilde{\mathbf{b}}_j   \right)^2\\
& =  \left[ s+ \gamma_{RL}- \sum_j \left(\mathbf{b}_j - i\tilde{\mathbf{b}}_j\right) \right]\left[ s+ \gamma_{RL}- \sum_j \left(\mathbf{b}_j + i\tilde{\mathbf{b}}_j \right)\right].
\end{aligned}
\end{equation}
In particular, for the special case that $\sin\left(\omega_a\tau_j\right) = 0$, we get  $\tilde{\mathbf{b}}_j = 0$, and consequently $\mathbf{x}(t)$ is exponentially stable when the time delays are small, which agrees with the conclusion in \textbf{Theorems~\ref{chiralsteady}-\ref{Promatain}}. On the other hand, when $\sin\left(\omega_a\tau_j\right) \neq 0$, the real parts of the roots of Eq.~(\ref{con:DeltasSCase1}) are negative, and thus $\mathbf{x}(t)$  converges to zero too.
\end{sloppypar}

Eq.~(\ref{con:linearControlxtLast}) can be rewritten as the following quasi-polynomial~\citep{angulo2019quasi}
\begin{small}
\begin{equation} \label{con:linearControlxtV2}
\begin{aligned}
&\Delta_{c} (s) =\left( s+ \gamma_{RL}- \sum_j \mathbf{b}_je^{-\tau_j s}\right)^2 +\left(\sum_j \tilde{\mathbf{b}}_j e^{-\tau_j s}  \right)^2\\
& = \left( s+ \gamma_{RL} \right)^2 - 2\sum_j \mathbf{b}_j\left( s+ \gamma_{RL} \right) e^{-\tau_j s}+ \left(\sum_j \mathbf{b}_je^{-\tau_j s} \right)^2
+ \left(\sum_j \tilde{\mathbf{b}}_j e^{-\tau_j s}  \right)^2\\
& = \sum_{q = 0}^2 \sum_{p=0}^5 a_{qp} s^q e^{-\Gamma_p s}, 
\end{aligned}
\end{equation}
\end{small}%
where $\Gamma_0 = 0$, $\Gamma_1 = \tau_1$, $\Gamma_2 = \tau_2$, $\Gamma_3 = 2\tau_1$, $\Gamma_4 = \tau_1 + \tau_2$, $\Gamma_5 = 2\tau_2$. The parameters $a_{qp}$ in Eq.~(\ref{con:linearControlxtV2}) are
$a_{20} = 1$, $a_{10} = 2\gamma_{RL}$, $a_{00} =\gamma_{RL}^2$,  and we can similarly derive the values of $a_{qp}$ for $p>0$. For simplicity, we denote 
\begin{equation} \label{con:linearControlxtV3}
\begin{aligned}
&\Delta_{c} (s) = \sum_{p=0}^5 L_p(s) e^{-\Gamma_p s}, 
\end{aligned}
\end{equation}
with $L_p(s) =\sum_{q = 0}^2  a_{qp} s^q$.
Because $\tau_2 \geq \tau_1$, $0\leq \Gamma_p/\Gamma_5 \leq \left(\tau_1+\tau_2 \right)/2\tau_2\leq 1$ for $p =0,1,2,3,4$. The quasi-polynomials in Eqs.~(\ref{con:linearControlxtV2},\ref{con:linearControlxtV3}) will be used to analyze the quantum coherent feedback control dynamics, in combination with the master equation given bellow.

\subsubsection{Master equation representation}  
Within the Markovian approximation, the transmission delay in the waveguide can be interpreted as a delayed phase, and the quantum state dynamics is governed by the master equation~\citep{Anton2018PRL,soro2022chiral}
\begin{equation} \label{con:TwoatomMasterEQ}
\begin{aligned}
\dot{\rho}(t) 
&=  -i\left [\sum_{j=1,2}H_{\mathrm{eff}}^{(j)} + H_{\mathrm{eff}}^{\rm I}
,\rho(t) \right]  + \sum_{j=1,2}\tilde{\Gamma}_{j} \mathcal{L}_{j} \left [ \rho(t) \right ] +\mathcal{D}_{12}\left [ \rho(t) \right ].
\end{aligned}
\end{equation}
Here, the Hamiltonian terms are
\begin{equation} 
\begin{aligned}
H_{\mathrm{eff}}^{(j)} =- \gamma_{jL}\gamma_{jR}\sin\left(\omega_a \frac{2z_j}{c}\right)\sigma_j^+  \sigma_j^-, 
\end{aligned}
\end{equation}
\begin{equation} 
\begin{aligned}
&H_{\mathrm{eff}}^{\rm I} =   \left(\frac{\gamma_{1L}\gamma_{2L}}{4}+\frac{\gamma_{1R}\gamma_{2R}}{4}\right)\sin\left(\omega_a \frac{z_2-z_1}{c}\right)\sigma_1^+  \sigma_2^- \\
&-\left(\frac{\gamma_{1L}\gamma_{2R}}{4}+\frac{\gamma_{1R}\gamma_{2L}}{4}\right)\sin\left(\omega_a \frac{z_1+z_2}{c}\right)\sigma_1^+  \sigma_2^-  + \rm{H.c.},
\end{aligned}
\end{equation}
the Lindblad components are
\begin{equation} 
\begin{aligned}
\mathcal{L}_j  \left [ \rho(t) \right ] = \sigma_j^- \rho(t)\sigma_j^+ -\frac{1}{2} \rho(t)\sigma_j^+\sigma_j^- - \frac{1}{2}\sigma_j^+  \sigma_j^-\rho(t)
\end{aligned}
\end{equation}
with coefficient \citep{Anton2018PRL}
\begin{equation} \label{gammajdecayrate}
\begin{aligned}
\tilde{\Gamma}_{j} =\frac{\gamma_{jR}^2+\gamma_{jL}^2}{2}  -\gamma_{jL}\gamma_{jR}\cos\left(\omega_a \frac{2z_j}{c} \right),
\end{aligned}
\end{equation}
and
\begin{equation} 
\begin{aligned}
&\mathcal{D}_{12}\left [ \rho(t) \right ]\\
=&\Gamma_{\rm coll} \left[\sigma_1^- \rho(t)\sigma_2^+ -\frac{1}{2} \rho(t)\sigma_1^+\sigma_2^- - \frac{1}{2}\sigma_1^+  \sigma_2^-\rho(t) +\rm{H.c.}\right],
\end{aligned}
\end{equation}
which represents the collective relaxation process of the two atoms \citep{Anton2018PRL}, where
\begin{equation}  \label{eq:Gamma_coll}
\begin{aligned}
\Gamma_{\rm coll} = &\left(\gamma_{1R}\gamma_{2R}+\gamma_{1L}\gamma_{2L}\right) \cos\left(\omega_a \frac{z_2-z_1}{c}\right)\\
&-\left(\gamma_{1L}\gamma_{2R}+\gamma_{1R}\gamma_{2L}\right) \cos\left(\omega_a \frac{z_1+z_2}{c}\right).
\end{aligned}
\end{equation}
In particular, when $\gamma_{jL} = \gamma_{jR}$ for $j=1,2$, the  master equation \eqref{con:TwoatomMasterEQ} reduces to the non-chiral case in~\citep{Anton2018PRL}.
\begin{remark}
Because initially the two atoms are both excited, in our case it is possible that $\rm{Rank} \left[\rho(t)\right] = 4$ when $t>0$, while in \citep{ZhangBin} for the single-excitation case,  $\rm{Rank}\left[\rho(t)\right] \leq 3$.
\end{remark}

\subsection{Atomic dynamics and photon emission analysis}
In this subsection, we analyze the atomic dynamics based on the quasi-polynomial model and master equation presented in the preceding subsections. 
\begin{Theorem} \label{Smalleststability}
The smallest value  of the $j$th atom's independent decay rate $\tilde{\Gamma}_{j}$ is attained when $z_j = n\pi/\omega_a$,  where $n=1,2,\cdots$.
\end{Theorem}
\begin{Proof}
The $j$th atom's decay rate $\tilde{\Gamma}_{j}$ to the waveguide is  given by Eq.~(\ref{gammajdecayrate}). Clearly, its smallest value for the given $\gamma_{jR}$ and $\gamma_{jL}$ is reached when $z_j = n\pi/\omega_a$.
\end{Proof}
According to \textbf{Theorem~\ref{Smalleststability}}, if the amplitudes for the atom's excited state  exponentially converges to zero when $z_j = n\pi/\omega_a$, then it will exponentially converge to zero for arbitrary delays. Thus, it suffices to study the stability when $z_j = n\pi/\omega_a$. In this parameter setting, the control equation of $c_{ee}(t)$ reduces to a real-value equation as $c_{ee}^I(t) \equiv 0$ in Eq.~(\ref{con:CeeRealImag}).

When $z_j = n\pi/\omega_a$ and $z_2/z_1 = m$, where $m$ is an integer, Eq.~(\ref{ceeRdot}) is simplified as
\begin{equation} \label{con:ceerealsimple}
\begin{aligned}
\dot{c}_{ee}^R(t) =&-\gamma_{RL}c_{ee}^R(t) + \mathbf{g}_1c_{ee}^R\left(t-\tau_1\right)\cos\left(\omega_a\tau_1\right) \\
&+ \mathbf{g}_2 c_{ee}^R\left(t-\tau_2\right)\cos\left(\omega_a\tau_2\right).
\end{aligned}
\end{equation}
Denote $\tilde{z} = e^{-z_1s}$, and define the quasi-polynomial function $F\left(\tilde{z}\right) = \sum_{p=0}^2 F_p \tilde{z}^p$, according to the general representation in Eq.~(\ref{con:linearControlxtV2}).

\begin{remark} \label{Largedelay}
As in Eq.~(\ref{con:ceerealsimple}), the quantum state dynamics can be influenced by time delays $\tau_j$. Compared with atom's slowest decaying to the waveguide independent from the coherent feedback loop, namely $\Gamma_0 = \min_{j}\left\{\left(\gamma_{jR}^2+\gamma_{jL}^2\right)/2\right\} $ according to Eq.~(\ref{gammajdecayrate}), we denote the large delay circumstance for the parameter setting that $\tau_j\gg 1/\Gamma_0$.
\end{remark}

Based on \textbf{Theorem~\ref{Smalleststability}}, the following \textbf{Theorem~\ref{kamen}} gives the condition when the systems is delay-independent stable.
\begin{Theorem} \label{kamen}
\citep{kamen1980relationship,kamen1982linear} When $z_2/z_1 = m$, where $m$ is an integer, Eq.~(\ref{ceeRdot}) is delay-independent stable if
\begin{description}
    \item[1)] the quasi-polynomial \[
P\left(s,e^{i\omega}\right) \neq 0, ~~ \rm{Re} (s) \geq 0, \omega \in [0,2\pi],
\]
where $P\left(s,e^{-\tau_1s}\right)$ is the quasi-polynomial defined according to Eq.~(\ref{ceeRdot}),
or
 \item[2)]    the real part of all the eigenvalues of $F\left(\tilde{z}\right)$ is negative.
\end{description}
\end{Theorem}

Furthermore, the stability of the coherent feedback control network when the atoms are chirally coupled to the waveguide is illustrated by the following theorem. 
\begin{Theorem} \label{Theorem6Chiral} 
When the couplings between atoms and waveguide are chiral, the coherent feedback network is exponentially stable independent of delay.
\end{Theorem}
\begin{Proof} 
In the master equation (\ref{con:TwoatomMasterEQ}), when the atoms are chirally coupled to the waveguide, the amplitudes of the Lindblad components $\tilde{\Gamma}_{j}  >0$  no matter whether the delays are small or large due to \textbf{Remark~\ref{Largedelay}}, which means that the atoms  exponentially decay to the ground states, independent of delays.  
\end{Proof}

 Based on Eq.~(\ref{con:TwoatomMasterEQ}), we have the following result on the spontaneous emission rate of the two-atom network.
\begin{Theorem}\label{CollectiveRelaxation} 
When $\gamma_{1R} = \gamma_{1L}$, $z_1= n\pi/\omega_a \ll c$ for some positive integer $n$, there is no collective relaxation between the two atoms, and instead the two atoms emit the coherent field independently to the waveguide.
\end{Theorem}
\begin{Proof}
The proof is straightforward because $\Gamma_{\rm coll} = 0$ in Eq. \eqref{eq:Gamma_coll} when the condition is satisfied.
\end{Proof}
\begin{remark}
A special case of  \textbf{Theorem}~\ref{CollectiveRelaxation} is \textbf{Theorem}~\ref{onephoton} for the generation of single-photon state in the waveguide. As there is no collective relaxation between the two atoms, the second atom at $z_2$ emits one photon into the waveguide and decays to its ground state due to its chiral coupling to the waveguide, while the first atom at $z_1$ remains excited.
\end{remark}

To conclude, because there are no external drives applied upon the quantum coherent network, the stability evaluated by the convergence rate of the quantum state amplitudes is related to the spontaneous emission rate of the atoms. This can be equivalently analyzed based on Eq.~(\ref{con:Ceetapprox}) and Eq.~(\ref{con:TwoatomMasterEQ}), as detailed below.

On one hand, the two-photon emission rate is determined by the localization of the poles in Eq.~(\ref{eq:sep23_C_ee}), or the time-domain representation of the amplitudes that two atoms are excited in  Eq.~(\ref{con:Ceetapprox}).  When the delay is small, the spontaneous emission rate can be approximated as 
\[
\gamma_{RL} - \gamma_{1L}\gamma_{1R} \cos\left(2\omega_a z_1/c\right) - \gamma_{2L}\gamma_{2R}\cos\left(2\omega_a z_2/c\right)\geq 0,
\]
according to Eq.~(\ref{con:Ceetapprox}). When the delay is large, i.e., $2z_1/c \gg 1/\Gamma_0$ due to \textbf{Remark~\ref{Largedelay}}, the spontaneous emission rate is $\gamma_{RL} $ when $t<2z_1/c$, and can be re-excited by absorbing the photon in the waveguide when $t>2z_1/c$.

On the other hand, when the evolution of quantum states is modeled with the master equation (\ref{con:TwoatomMasterEQ}), and the waveguide is regarded as an environment, the spontaneous emission rate of atoms can be evaluated by the amplitudes of the Lindblad component, namely $\tilde{\Gamma}_{j}$ in Eq.~(\ref{gammajdecayrate}), as analyzed in \textbf{Theorems~\ref{Smalleststability}} and \textbf{\ref{Theorem6Chiral}}.

\section{Coherent feedback control analyzed in the spatial domain} \label{Sec:position}
\begin{sloppypar}
The number of photons in the waveguide can be studied by means of delay-dependent feedback equations (\ref{con:Ceedelay}-\ref{con:Cgedelay}). However, the spatial distribution of photons in the waveguide can only be studied by modeling in the spatial domain. The waveguide Hamiltonian in Eq.~(\ref{con:H0}) can be equivalently represented in spatial domain as~\citep{Shentheory1,boundary}
\begin{small}
\begin{equation} \label{con:HamWavePos}
H_w  = i v_g \int_{0}^{\infty}  c_L^{\dag}(z) \frac{\partial}{\partial z} c_L(z)\mathrm{d}z  - i v_g\int_{0}^{\infty}  c_R^{\dag}(z) \frac{\partial}{\partial z} c_R(z)\mathrm{d}z,
\end{equation}
\end{small}%
where  $v_g$ is the group velocity of the photonic  wavepacket in the waveguide,  $c_L^{\dag}(z)$ and $c_L(z)$ are the creation and annihilation operators for the left-propagating field at the position $z$, and  $c_R^{\dag}(z)$ and $c_R(z)$ are those for the right-propagating field. The derivation of  Eq. \eqref{con:HamWavePos} is further introduced in \textbf{~\ref{Sec:PositionHam}}. Based on this, we study the coherent feedback dynamics with one or two atoms as follows.
\end{sloppypar}

\subsection{One atom coupled to the waveguide} \label{subsec:one atom}
When there is only the first atom at $z=z_1$ in Fig.~\ref{fig:scheme} coupled to the waveguide, namely $\gamma_{2L} = \gamma_{2R} = 0$, the quantum state with one exciation is
\begin{small}
\begin{equation} \label{con:statepacketInteraction}
\begin{aligned}
|\Psi(t)\rangle =& \int_0^{\infty}  \Phi_R(z,t)e^{-i\tilde{\omega}_1 t} |1_{z}^r,g\rangle \mathrm{d}z \\
&+ \int_0^{\infty} \left[ \Phi_L(z,t) e^{-i\tilde{\omega}_1 t} |1_{z}^l,g\rangle + c_{e}(t)e^{-i\tilde{\omega}_2 t}|0,e\rangle \right]
\mathrm{d}z,
\end{aligned}
\end{equation}
\end{small}%
where $|1_{z}^{r,l},g\rangle$ represents that the first atom is at the ground state and the right or left-propagating photon can be detected at the position $z$, $|0,e\rangle$ means that the first atom is excited and the waveguide is empty, $\Phi_{R}(z,t)$ denotes the amplitude of the right-propagating photon wavepacket at the position $z$, $\Phi_{L}(z,t)$ is the amplitude of a  left-propagating photon wavepacket, $c_e(t)$ is the amplitude that the first atom is excited, and $\tilde{\omega}_1$ and $\tilde{\omega}_2$ are the eigenfrequencies of the ground and excited state respectively with $\omega_a = \tilde{\omega}_2 -\tilde{\omega}_1$.

The Hamiltonian of the system reads
\begin{equation} \label{con:HamiltonainZdomMarkoviaChiral0}
H  =\tilde{\omega}_1 |g\rangle \langle g| +\tilde{\omega}_2 |e\rangle \langle e| +H_w +H_m + H_{\rm I},
\end{equation}
where the waveguide Hamiltonian $H_w$ is given by Eq. \eqref{con:HamWavePos}, the mirror Hamiltonian $H_m$ is given by Eq.~\eqref{con:HamiltonainMirror1} as \textbf{Lemma \ref{lemmaMirror}} in \textbf{\ref{Sec:PositionHam}}, and the interaction Hamiltonian $H_{\rm I}$ can be derived by transforming its format in the frequency domain (as in Eq.~(\ref{con:Hintj})) to the spatial domain, which is Eq.\eqref{con:HIposition} in \textbf{\ref{Sec:PositionHam}}.

Solving the Schr\"{o}dinger equation $\frac{\mathrm{d}}{\mathrm{d} t}|\Psi(t)\rangle = -i H |\Psi(t)\rangle$ with $H$ in Eq. \eqref{con:HamiltonainZdomMarkoviaChiral0} and the ansatz $|\Psi(t)\rangle$ in Eq. \eqref{con:statepacketInteraction}, we can derive the partial differential equations (PDEs) as
\begin{subequations} \label{con:chiralOneatomODEFinal}
\begin{numcases}{}
\dot{c}_e(t) = -\left[\gamma_{1R} \Phi_R(z_1,t)  + \gamma_{1L}\Phi_L(z_1,t) \right]e^{i\omega_a t},\label{Posmodel1}\\
\frac{\partial \Phi_R(z,t)}{\partial t} = -v_g\frac{\partial \Phi_R(z,t)}{\partial z} + 2v_g\delta(z)  \Phi_L(z,t)e^{i2kz} \notag\\
~~~~~~~~~~~~~~~~~~~+ \gamma_{1R}\delta(z-z_1)c_e(t)e^{-i\omega_a t},\label{Posmodel2}\\
\frac{\partial \Phi_L(z,t)}{\partial t} =  v_g \frac{\partial \Phi_L(z,t)}{\partial z}  - 2v_g\delta(z)  \Phi_R(z,t)e^{-2ikz} \notag\\
~~~~~~~~~~~~~~~~~~~+ \gamma_{1L}\delta(z-z_1)c_e(t)e^{-i\omega_a t} ,\label{Posmodel3}
\end{numcases}
\end{subequations}
where Eq.~(\ref{Posmodel1}) represents that an excited atom can emit photon wave packet into the waveguide along the right or left direction, and the distribution of photon wave packet is determined by the chiral coupling strengths between the atom and waveguide; Eq.~(\ref{Posmodel2}) shows that the right-propagating mode is determined by mirror's reflection of a left-propagating packet at $z=0$, the atom's position, and the coupling strength between the atom and the waveguide mode along the right direction; Eq.~(\ref{Posmodel3}) further shows that the left-propagating mode can be influenced by the coupling strength between the atom and the waveguide along the left direction.

\begin{remark}
One of the advantages of Eq.~(\ref{con:chiralOneatomODEFinal}) in the spatial domain is that the differential of the excited atomic state is directly represented with the spatial distribution of the photon states rather than their integrations as in Eq.~(\ref{con:Popuquation}).
\end{remark}

The quantum state with one right- or left-propagating photon in the waveguide $\Phi_R(z,t)$ and $\Phi_L(z,t)$ in Eq. \eqref{con:statepacketInteraction} can be represented according to the position $z$ as
\begin{subequations} \label{con:PositionfunctionStep}
\begin{numcases}{}
\Phi_R(z,t) = \left[\Theta(z) -\Theta\left(z-z_1\right)\right] f_r\left(t-\frac{z}{v_g}\right) \notag\\
~~~~~~~~~~~~~~~+  \Theta\left(z-z_1\right) g_r\left(t-\frac{z}{v_g}\right), \label{con:PositionfunctionStep_a}\\
\Phi_L(z,t) = [\Theta(z) -\Theta\left(z-z_1\right)] f_l\left(t+z/v_g\right),
\label{con:PositionfunctionStep_b}
\end{numcases}
\end{subequations}
respectively, where $\Theta$ represents the Heaviside step function, $f_r\left(t-z/v_g\right)$ represents the right-propagating photonic mode in the area $0<z<z_1$,  $g_r\left(t-z/v_g\right)$ is that in the area $z>z_1$, and $f_l\left(t+z/v_g\right)$ represents the left-propagating photonic mode in the area $0<z<z_1$~\citep{boundary}.

According to the calculations in \textbf{\ref{Sec:oneatomApendix}}, Eq.~(\ref{Posmodel1}) can be written in the delay-dependent form as
\begin{equation} \label{con:MaintexCedelayCal}
\begin{aligned}
\dot{c}_e(t)=-\frac{\gamma_{1R}^2 + \gamma_{1L}^2}{2v_g} c_e(t) +\frac{\gamma_{1L}\gamma_{1R}}{v_g}c_e\left(t-\frac{2z_1}{v_g}\right)e^{i\frac{2\omega_a z_1}{v_g}}.
\end{aligned}
\end{equation}

Let the atom be initially excited, then Laplace transforming Eq. \eqref{con:MaintexCedelayCal} yields
\begin{equation} \label{con:CedelayCalPosFinal}
\begin{aligned}
C_e(s) &= \frac{1}{s-\gamma_{1L}\gamma_{1R}e^{-2z_1s/v_g}e^{i2\omega_a z_1/v_g} +\frac{\gamma_{1R}^2 + \gamma_{1L}^2}{2}},
\end{aligned}
\end{equation}
which agrees with Eq.~(\ref{con:Ceedelay}) with $\gamma_{2R} = \gamma_{2L} = 0$. 

\begin{Theorem} \label{OneAtomPosition}
When $\gamma_{1R} = \gamma_{1L}$, $z_1 =  n\pi/\omega_a$, where $n$ is an integer satisfying that $n\pi/\omega_a \ll v_g$, the atom at $z_1$ can be sustainably  excited.
\end{Theorem}

\begin{Proof}
\begin{sloppypar}
When $\gamma_{1R} = \gamma_{1L}$, and $z_1 =  n\pi/\omega_a \ll v_g$, $-\gamma_{1L}\gamma_{1R}e^{-2z_1s/v_g}e^{i\omega_a \tau_1} +\left(\gamma_{1R}^2 + \gamma_{1L}^2\right)/2 \approx \left(\gamma_{1R}^2 + \gamma_{1L}^2\right)/2 -\gamma_{1L}\gamma_{1R} = 0$. Then
\end{sloppypar}
\begin{equation} \label{con:CesAppro}
\begin{aligned}
C_e(s) \approx \frac{1}{s-\gamma_{1L}\gamma_{1R} +\frac{\gamma_{1R}^2 + \gamma_{1L}^2}{2}} = \frac{1}{s},
\end{aligned}
\end{equation}
and $\lim_{t\rightarrow \infty} c_e(t) = \lim_{s\rightarrow 0} sC_e(s) = 1$.
\qed
\end{Proof}

\begin{figure}[h]
\centerline{\includegraphics[width=0.7\columnwidth]{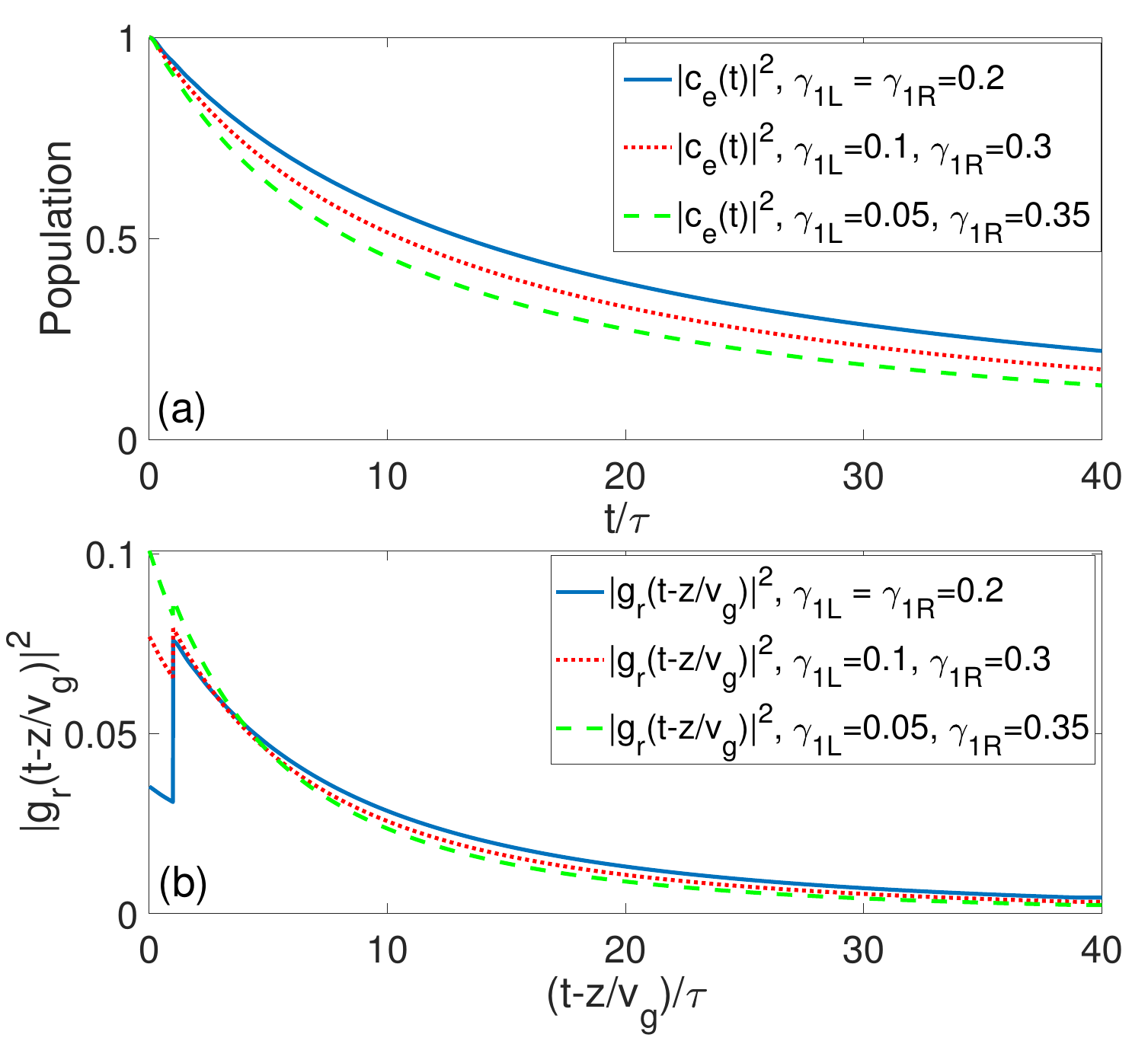}}
\caption{Comparisons on the atom's excited populations (a) and right-propagating photon wave packet (b) influenced by the coherent feedback control relied on chiral and nonchiral couplings between one atom and the semi-infinite waveguide.}
	\label{fig:OneatomCompare}
\end{figure}

\begin{sloppypar}
We generalized the control performance that the atom is nonchirally coupled to the waveguide in ~\citep{boundary} to that the atom is chirally coupled to the waveguide, rendering directional propagating of photon wave packets. Compared with~\citep{boundary}, where the mirror is at the right terminal of the waveguide, we first derive its counterpart as Eq.~\eqref{con:HamiltonainMirror1} when the mirror is at the left terminal of the waveguide. After that we compare the populations of the excited atomic state and the right propagating single-photon wave packet in Fig.~\ref{fig:OneatomCompare}, where (a) is for the population $|c_e(t)|^2$ and (b) is for the right-propagating photon wave packet solved in \textbf{\ref{Sec:oneatomApendix}}. For the simulations of the right-propagating photon wave packet, $g_r\left(t-z/v_g\right)$ can be derived by replacing $t$ in the right hand side of Eq.~(\ref{con:grsolution}) with $t-z/v_g$. 
\end{sloppypar}

As for the parameters settings in Fig.~\ref{fig:OneatomCompare}, $\omega_a = 50$, $z_1 = 2.25\pi/\omega_a$ and $\gamma_{1L} + \gamma_{1R} = 0.4$, then $|c_e(t)|^2= e^{-(\gamma_{1L}^2 + \gamma_{1R}^2)t} \leq e^{-(\gamma_{1L}+\gamma_{1R})^2 t/2}$ with the equality holds only when $\gamma_{1L} = \gamma_{1R}$. Thus the chiral coupling between one atom and waveguide can induce faster decaying of the excited atomic state. 

The conclusion in \textbf{Theorem~\ref{OneAtomPosition}} agrees with the analysis in the frequency domain in ~\citep{waveguideOneatom}, which clarifies that at the exact parameter setting of the first atom's position, the atom's excited population can be unchanged rather than converge to zero. Considering that \textbf{Theorem~\ref{OneAtomPosition}} holds only when the interaction between the atom and the waveguide is nonchiral, this can also be derived from the nonchiral model in ~\citep{boundary}.

\subsubsection{The influence of the feedback loop length on stability}
\begin{sloppypar}
The stability of linear system with time delays according to quasi-polynomial approach~\citep{malakhovski2006stability,kharitonov2002robust}, and the stability can be different due to the value of time delays~\citep{TAC1984olbrot}. As illustrated by \citep{TAC1984olbrot}, when $\gamma_{1L} \neq \gamma_{1R}$ or $\cos\left(\omega_a \tau_1\right) <1$, the linear system with time delay (\ref{con:MaintexCedelayCal}) is asymptotically stable. However, when the delay $\tau_1$ is large due to \textbf{Remark~\ref{Largedelay}}, the system may not be exponentially stable. To clarify this, we first rewrite the delay-dependent equation (\ref{con:MaintexCedelayCal}) as real-value equations in terms of the real and imaginary parts of $c_e(t)$, which is
\end{sloppypar}
\begin{subequations} \label{con:RICe}
\begin{align}
&\dot{c}_e^R(t) = -\gamma_{RL}^{(1)} c_e^R(t) \notag\\
&~~~~~+ \mathbf{g}_1 \left[ c_e^R\left(t-\tau_1\right)\cos\left(\omega_a \tau_1\right) - c_e^I\left(t-\tau_1\right)\sin\left(\omega_a \tau_1\right)\right],\\
&\dot{c}_e^I(t) = -\gamma_{RL}^{(1)} c_e^I(t) \notag\\
&~~~~~+ \mathbf{g}_1 \left[ c_e^R\left(t-\tau_1\right)\sin\left(\omega_a \tau_1\right) + c_e^I\left(t-\tau_1\right)\cos\left(\omega_a \tau_1\right)\right] ,
\end{align}
\end{subequations}
where $c_e(t) = c_e^R(t) + ic_e^I(t)$, $\gamma_{RL}^{(1)} = \left(\gamma_{1R}^2 + \gamma_{1L}^2\right)/2 $. Denote $\mathbf{x}_1(t) = \left[c_{e}^R(t),c_{e}^I(t) \right]^T$, $\tilde{A}= {\rm diag}\left(-\gamma_{RL}^{(1)}  ,-\gamma_{RL}^{(1)} \right) $, and\\ $\tilde{B} = \mathbf{g}_1 \begin{bmatrix}
   \cos\left(\omega_a \tau_1\right)  & -\sin\left(\omega_a \tau_1\right) \\
    \sin\left(\omega_a \tau_1\right) &\cos\left(\omega_a \tau_1\right) \\
  \end{bmatrix}$. Then Eq. \eqref{con:RICe} becomes
  \begin{equation} \label{con:OneAtomVecEq}
\begin{aligned}
\dot{\mathbf{x}}_1(t) =\tilde{A}\mathbf{x}_1(t) + \tilde{B} \mathbf{x}_1\left(t-\tau_1\right),
\end{aligned}
\end{equation}
whose characteristic quasi-polynomial is
\begin{equation} \label{con:Delta1S}
\begin{aligned}
&~~~~\Delta_{1} (s) = \left | sI -\tilde{A} -\tilde{B}  e^{-\tau_1 s} \right|\\
& = \left(s+ \gamma_{RL}^{(1)} \right)^2 -2\mathbf{g}_1\cos\left(\omega_a \tau_1\right) \left(s+ \gamma_{RL}^{(1)} \right)e^{-\tau_1 s} + \mathbf{g}_1^2 e^{-2\tau_1 s}.
\end{aligned}
\end{equation}
Next, we consider the circumstance that $\tau_1$ is large according to \textbf{Remark~\ref{Largedelay}}.  
We denote the solutions of $\Delta_{1} (s) = 0$ as $\tilde{s}$ for convenience.

\begin{Theorem} \label{LargedelayOneatomEps}
For arbitrary $\epsilon > 0$, there exists $\tau_1>\tau_{\epsilon}>0$, such that there is one root of Eq.~(\ref{con:Delta1S}) which satisfies that $\rm Re\left(\tilde{s}\right) \geq -\epsilon$ and therefore the atomic steady state is not the ground state.
\end{Theorem}
\begin{Proof}
The proof of $\rm Re\left(\tilde{s}\right) \geq -\epsilon$ is similar to that of Theorem 1 in \citep{TAC1984olbrot}, thus is omitted. This, together with that there are no oscillating behavior as determined by the Lindblad master equation~(\ref{con:TwoatomMasterEQ}),  the steady value of $c_e(t)$ can be nonzero, which means the atomic steady state is not the ground state.
\end{Proof}

\begin{figure}[h]
\centerline{\includegraphics[width=1\columnwidth]{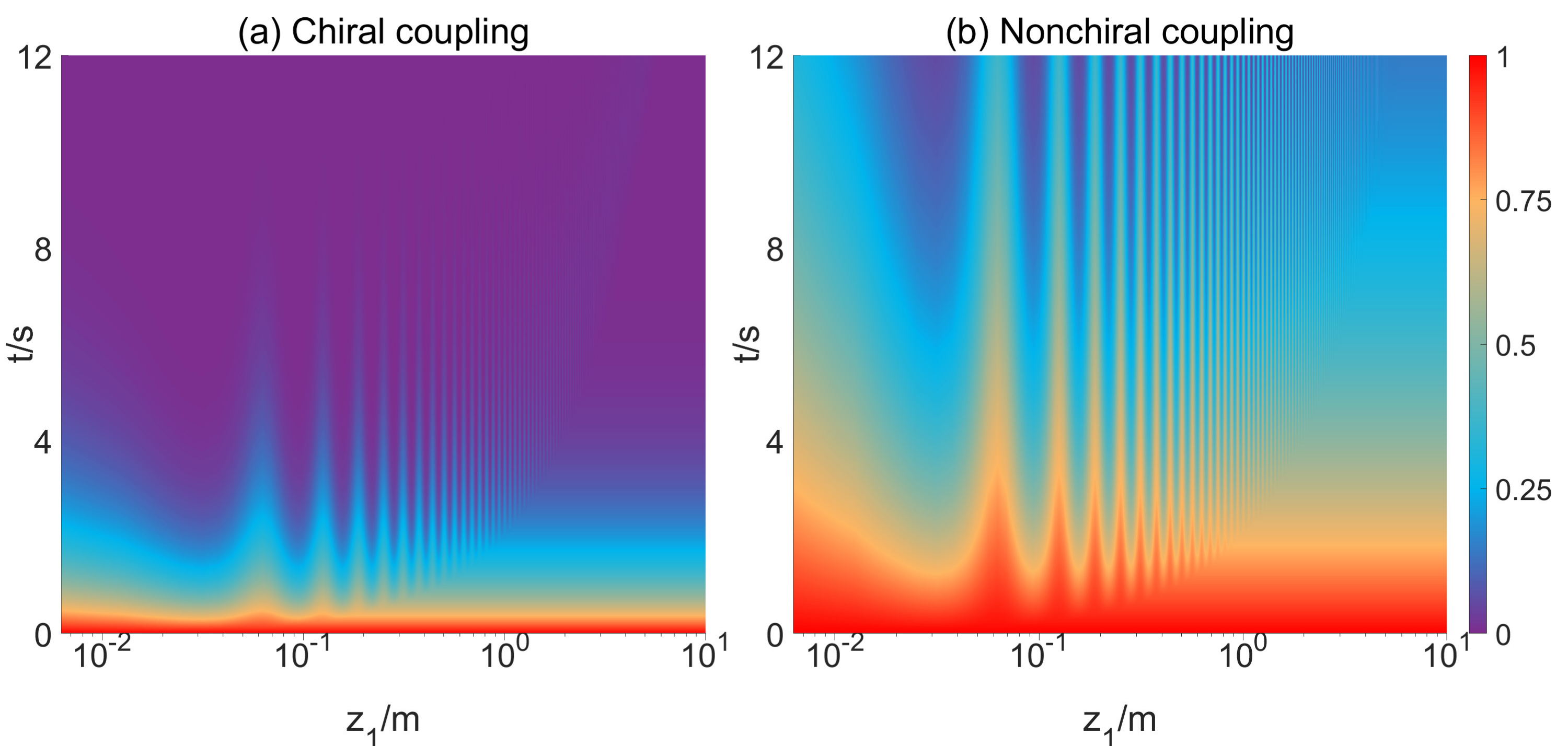}}
\caption{Atom populations influenced by time delays when the atom is chirally (a) or nonchirally (b) coupled to the waveguide.}
	\label{fig:largedelayOscilate}
\end{figure}

Based on \textbf{Theorem~\ref{LargedelayOneatomEps}}, the influence of atomic dynamics by time delays is further illustrated in the following simulations. In Fig.~\ref{fig:largedelayOscilate}, one two-level atom is coupled to the semi-infinite waveguide at $z_1$. Parameters are  $\omega_a = 50$, $\gamma_{1L} = 0.2$ and $\gamma_{1R} = 0.6$ in (a), and $\gamma_{1L} = \gamma_{1R}=0.2$ in (b). As shown in the chiral parameter setting in (a), an initially excited atom  certainly decays to the ground state. Moreover, the decaying rate is much smaller when $z_1 = n\pi/\omega_a$ with $n=1,2,\cdots$, which agrees with Eq.~(\ref{con:CedelayCalPosFinal}), \textbf{Theorem~\ref{OneAtomPosition}}, or the simplified  \textbf{Theorem~\ref{chiralsteady}}. On the other hand, in the nonchiral parameter setting in (b),  the large delay can induce non-zero steady excited populations,  which agrees with the conclusion in \textbf{Theorem~\ref{LargedelayOneatomEps}}.

\begin{remark}
When the atom is far from the terminal mirror of the semi-infinite waveguide, the emitted filed by the atom to the mirror can re-excite the atom after being reflected, resulting in the non-zero populations of the atomic states. This means that the non-Markovian quantum dynamics realized by a coherent feedback channel with a large delay due to \textbf{Remark~\ref{Largedelay}} can make a difference compared with the circumstance with a short delay. 
\end{remark}

\subsection{Two-atom dynamics in the spatial domain} \label{subsec: 2 atom}
In this subsection, we analyze in the spatial domain on the dynamics that two atoms are coupled to a semi-infinite waveguide. When there are two atoms coupled to the semi-infinite waveguide, the Hamiltonian for the two-atom case, as the counterpart of that in Eq.\eqref{con:HamiltonainZdomMarkoviaChiral0} for the single-atom case,  reads
\begin{small}
\begin{equation} \label{con:HamiltonainZdomMarkoviaChiral}
\begin{aligned}
H =&\sum_{j=1,2}(\hbar\tilde{\omega}_{1}|g_j\rangle\langle g_j| + \hbar\tilde{\omega}_{2}|e_j\rangle\langle e_j|)
+H_w +H_m + \sum_{j=1,2} H_{\rm I}^{(j)},
\end{aligned}
\end{equation}
\end{small}%
where $H_w$ and $H_m$ are given in Eq.~(\ref{con:HamiltonainZdomMarkoviaChiral0}), and $H_{\rm I}^{(j)}$ is generalized from Eq.~(\ref{con:HIposition}) as 
\begin{small}
\begin{equation} 
\begin{aligned}
H_{\rm I}^{(j)} &= -i  \int_{-\infty}^{\infty} \left[\gamma_{jR} c_R(z) \delta(z-z_j) +\gamma_{jL} c_L(z) \delta(z-z_j) \right] \sigma_j^+ \mathrm{d}z  + \mathrm{H.c.},  
\end{aligned}
\end{equation}
\end{small}%
where $j=1,2$.

We assume that initially only the first atom is excited, and there can be at most one photon in the waveguide. Then the quantum state can be represented as
\begin{small}
\begin{equation} \label{con:TwoatomOneExit}
\begin{aligned}
&|\Psi(t)\rangle= \sum_{j=1,2} c_{j}(t)e^{-i\left (\tilde{\omega}_1+\tilde{\omega}_2\right )t} \sigma_j^+ |g_1,g_2,\{0\}\rangle \\
&+ \int_0^{\infty} \Phi_{g}^{l}(t,z)e^{-i2\tilde{\omega}_1 t} |g_1,g_2,1_z^r\rangle \mathrm{d}z + \int_0^{\infty} \Phi_{g}^{l}(t,z)e^{-i2\tilde{\omega}_1 t} |g_1,g_2,1_z^l\rangle \mathrm{d}z,
\end{aligned}
\end{equation}
\end{small}%
where $c_j(t)$ represents the amplitude that the $j$th atom is excited and the waveguide is empty, $\Phi_{g}^{r}(t,z)$  and $\Phi_{g}^{l}(t,z)$ represent the amplitudes of the states that both of the two atoms are in the ground state and there is one right- or left-propagating photon in the waveguide respectively. Taking the state representation into the Schr\"{o}dinger equation yields
\begin{subequations} \label{con:TwoatomOneEx}
\begin{align}
\dot{c}_{j}(t) =&  - \gamma_{jR} \Phi_{g}^{r}(t,z_j) e^{i\omega_at} -   \gamma_{jL} \Phi_{g}^{l}(t,z_j) e^{i\omega_at},\label{PosTAmodel1}\\
\frac{\partial \Phi_{g}^{r}(t,z)}{\partial t} =& -v_g \frac{\partial \Phi_{g}^{r}(t,z)}{\partial z} + 2v_g \Phi_{g}^{l}(t,z) e^{2ikz} \delta(z)\notag\\ 
&+ \sum_{j=1,2} \gamma_{jR} c_j(t) \delta(z-z_j) e^{-i\omega_a t},\label{PosTAmodel2}\\
\frac{\partial \Phi_{g}^{l}(t,z)}{\partial t} =& v_g \frac{\partial \Phi_{g}^{l}(t,z)}{\partial z}- 2v_g \Phi_{g}^{r}(t,z) e^{-2ikz} \delta(z) \notag\\
&+ \sum_{j=1,2} \gamma_{jL} c_j(t) \delta(z-z_j) e^{-i\omega_a t}.\label{PosTAmodel3}
\end{align}
\end{subequations}
Then $\Phi_{g}^{r}(t,z)$ and $\Phi_{g}^{l}(t,z)$ can be further represented as
\begin{small}    
\begin{subequations} \label{con:TwoatomPositionfunctionStep}
\begin{numcases}{}
\Phi_g^r(z,t) = \left[\Theta(z) -\Theta(z-z_1)\right] f_r\left(t-z/v_g\right)\notag\\
+ \left [ \Theta(z-z_1) - \Theta(z-z_2) \right ] g_r\left(t-z/v_g\right) +  \Theta(z-z_2) h_r\left(t-z/v_g\right), \label{con:PositionTwoAOneE_a}\\
\Phi_g^l(z,t) = \left[\Theta(z) -\Theta(z-z_1)\right] f_l\left(t+z/v_g\right) \notag\\
~~~~~~~~~~~~~~~+ \left [ \Theta(z-z_1) - \Theta(z-z_2) \right ] g_l\left(t+z/v_g\right),
\label{con:PositionTwoAOneEStep_b}
\end{numcases}
\end{subequations}
\end{small}%
where $f_r\left(t-z/v_g\right)$, $g_r\left(t-z/v_g\right)$ and $h_r\left(t-z/v_g\right)$ represent the right-propagating photon wave packets at $[0,z_1]$, $[z_1,z_2]$ and $[z_2,+\infty]$, $f_l\left(t+z/v_g\right)$ and $g_l\left(t+z/v_g\right)$ are for the left-propagating photon wave packets at $[0,z_1]$ and $[z_1,z_2]$, respectively.

\begin{figure}[h]
\centerline{\includegraphics[width=0.7\columnwidth]{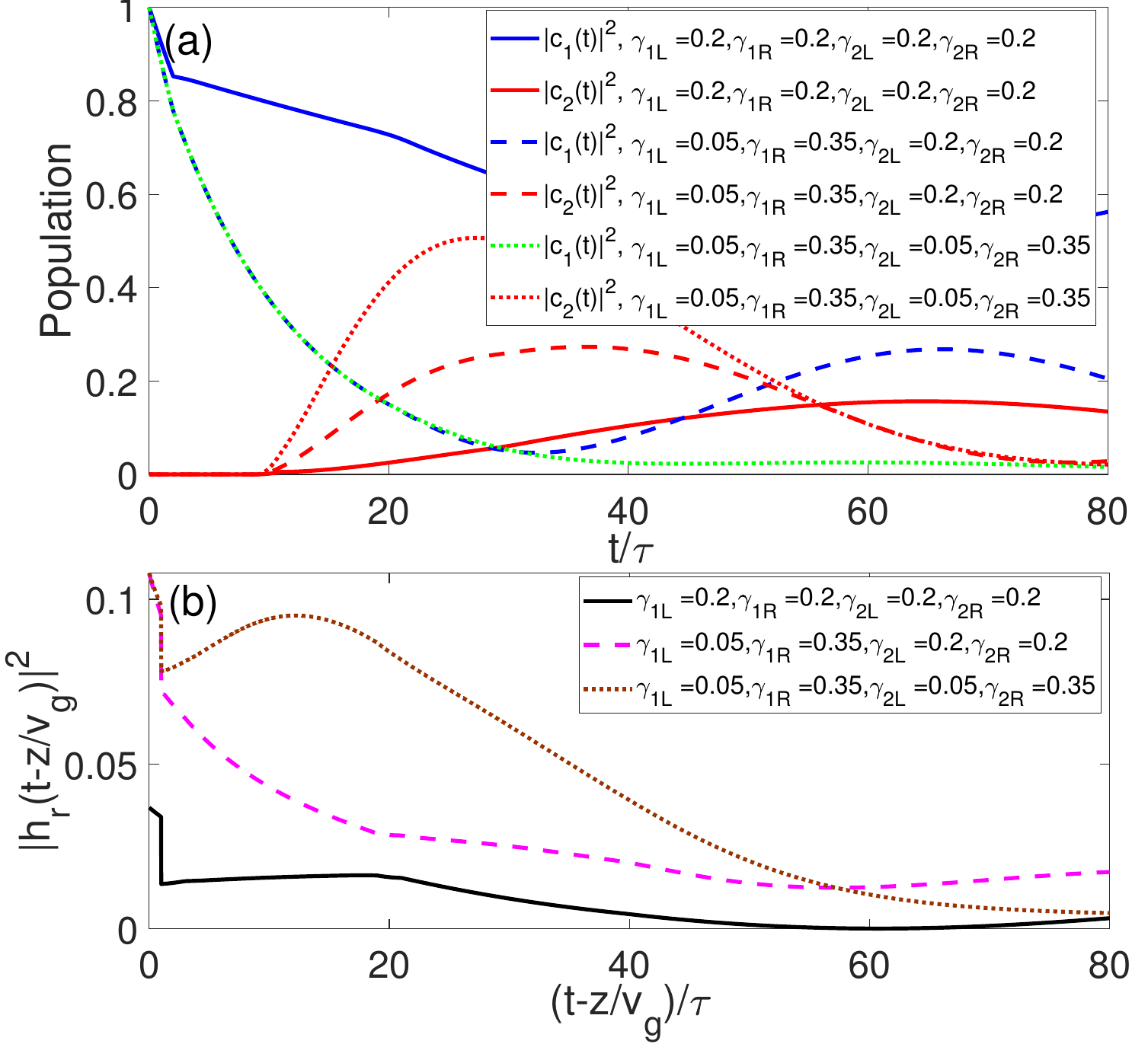}}
\caption{Comparison on the atoms' excited populations (a) and the right-propagating photon wave packet (b) of the coherent feedback network where two atoms are coupled to a semi-infinite waveguide with one excitation.}
	\label{fig:TwoatomOneExcitationCompare}
\end{figure}

Furthermore, $c_j(t)$ with $j=1,2$ can be solved in the spatial domain as illustrated in \textbf{\ref{Sec:TwoatomOneExApendix}}, which agrees with the analysis in the frequency domain in \citep{ZhangBin}. While the spatial distribution of the photon wave packet can be clarified based on Eq.~(\ref{con:TwoatomOneEx}) and solved as Eq.~(\ref{con:Packetsolution}), which cannot be covered in the frequency domain analysis. As compared in Fig.~\ref{fig:TwoatomOneExcitationCompare} (a) with $z_2 = 10z_1 =10$ and $\omega_a = 50$, the first atom decays via the spontaneous emission, and the emitted photon can further excite the second atom after the direct transmission or the reflection by the mirror. As a result, the second atom can be transiently more exited by the right-propagating photon wave packet emitted by the first atom, especially when $\gamma_{2R}$ is much larger than $\gamma_{2L}$. The right-propagating photon wave packets at $z>z_2$ are compared in Fig.~\ref{fig:TwoatomOneExcitationCompare} (b) by combining Eqs.~(\ref{con:EdD7c},\ref{con:EdD7d},\ref{con:EdD7e}), which is influenced by atoms' positions and the reflection by the mirror.

\begin{remark}
\begin{sloppypar}The spatial domain analysis makes it possible to solve the spatial distribution and propagation of photon wavepackets, which cannot be acquired by modeling in the frequency domain. However, the atom population dynamics derived in the spatial domain must be equivalent with that derived in the frequency domain, as in \citep{ZhangBin,huo2020absorption,guimond2017delayed,ZollerPRL}. \end{sloppypar}
\end{remark}

\subsection{Spontaneous emission rate in the spatial domain}
On one hand, when one atom is coupled to the waveguide, as in Eq.~(\ref{con:MaintexCedelayCal}), the atom's spontaneous emission rate can be evaluated by $\left(\gamma_{1R}^2 + \gamma_{1L}^2\right)/2v_g$. This can also be generalized to the two-atom case in the spatial domain such as in Eq.~(\ref{con:c1dotCal}).

On the other hand,  the emission rate of photons can also be evaluated in the spatial domain as in Eqs.~(\ref{con:flt},\ref{con:grsolution}) for one-atom case and Eq.~(\ref{con:Packetsolution}) for two-atom case, respectively. It can be seen from Eq.~(\ref{con:Packetsolution}) that the photon emission rate towards mirror (left-propagating modes) in the waveguide and the emission rate leaving the system (right-propagating modes) are determined by the chiral couplings and atoms' positions.

\subsection{Non-exponentially stable dynamics induced by large coherent feedback delays}
When there are two atoms coupled to the semi-infinite waveguide, and initially only the first atom is excited, Eq.~(\ref{PosTAmodel1}) can be equivalently written in a delay dependent format as~\citep{guimond2017delayed,ZhangBin,ZollerPRL}
\begin{small}    
\begin{subequations} 
\begin{align}
&\dot{c}_1(t)  = - \frac{\gamma_{1R}^2 + \gamma_{1L}^2 }{2}c_1(t)-\gamma_{1L} \gamma_{2L} c_2\left (t-\frac{z_2-z_1}{v_g}\right )  e^{i\omega_a\frac{z_2-z_1}{v_g}}\notag\\
&+\gamma_{1R} \gamma_{1L} c_1\left (t-\frac{2z_1}{v_g}\right )  e^{i\omega_a \frac{2z_1}{v_g}} +\gamma_{1R} \gamma_{2L} c_2\left (t-\frac{z_1+z_2}{v_g}\right )  e^{i\omega_a \frac{z_1+z_2}{v_g}} , \label{con:C1tOnephoton}\\
&\dot{c}_2(t) = - \frac{\gamma_{2R}^2 + \gamma_{2L}^2 }{2}c_2(t) -\gamma_{2R} \gamma_{1R}  c_1\left (t-\frac{z_2-z_1}{v_g}\right )e^{i\omega_a\frac{z_2-z_1}{v_g}} \notag\\
&+\gamma_{2R} \gamma_{2L} c_2\left (t-\frac{2z_2}{v_g}\right )  e^{i\omega_a \frac{2z_2}{v_g}} +\gamma_{2R} \gamma_{1L} c_1\left (t-\frac{z_1+z_2}{v_g}\right )  e^{i\omega_a \frac{z_1+z_2}{v_g}} , \label{con:C2tOnephoton}
\end{align}
\end{subequations}
\end{small}%
as derived in the spatial domain in \textbf{\ref{Sec:TwoatomOneExApendix}} or in the frequency domain~\citep{ZhangBin}. The dynamics when $z_1$ and $z_2$ are small has been studied by~\citep{ZhangBin}. Now, we study the non-exponential stability dynamics when the delay is relative large according to \textbf{Remark~\ref{Largedelay}}.
\begin{Theorem} \label{LargedelayOneatom}
When $z_1 \lesssim z_2 = n\pi/\omega_a$ for some integer $n\gg 1$ and $\tau_1 \gg 0$, $\gamma_{1R} = \gamma_{1L} = \gamma_1$, $\gamma_{2R} = \gamma_{2L} = \gamma_2$, the two atoms have non-zero steady population satisfying $\left| c_1/c_2\right| = \gamma_1/\gamma_2$.
\end{Theorem}
\begin{Proof} 
In this parameter setting, the evolution of the atomic amplitudes read
\begin{footnotesize}
\begin{equation} \label{con:c1c2dot}
\begin{bmatrix}
\dot{c}_1(t)  \\
\dot{c}_2(t) \\
  \end{bmatrix}= \begin{bmatrix}
   -\gamma_1^2  & -\gamma_1\gamma_2 \\
  -\gamma_1\gamma_2 &-\gamma_2^2 \\
  \end{bmatrix}\begin{bmatrix}
c_1(t)  \\
c_2(t) \\
  \end{bmatrix} + e^{i\omega_a \tau_1}\begin{bmatrix}
   \gamma_1^2  & \gamma_1\gamma_2 \\
  \gamma_1\gamma_2 &\gamma_2^2 \\
  \end{bmatrix}\begin{bmatrix}
c_1\left(t-\tau_1\right)  \\
c_2\left(t-\tau_1\right) \\
  \end{bmatrix}.
\end{equation}
\end{footnotesize}%
When $n\gg 1$, $c_1(t)$ and $c_2(t)$ can be solved using the method introduced in the proof of Theorem 1 in ~\citep{ding2023quantum}, thus is omitted here. When $n$ and $\tau_1$ are large, $c_1\left(\tau_1\right)\approx \frac{\gamma_1^2}{\gamma_1^2 + \gamma_2^2}$, and $c_2\left(\tau_1\right)\approx -\frac{\gamma_1\gamma_2}{\gamma_1^2 + \gamma_2^2}$. When $t>t_1$, according to the proof of Theorem 1 in \citep{ding2023quantum}, $c_1(t)$ and $c_2(t)$ are asymptotically stable around $c_1\left(\tau_1\right)$ and $c_2\left(\tau_1\right)$ respectively.
\end{Proof}

\begin{sloppypar}
\textbf{Theorem~\ref{LargedelayOneatom}} means that when the delay in the waveguide-atom network is large due to \textbf{Remark~\ref{Largedelay}}, and the coupling between the atom and waveguide is nonchiral, the atomic state can be non-exponentially stable with non-zero steady values. A simplified example is the one-atom case in Fig.~\ref{fig:largedelayOscilate}(b), where the atom does not settle to  the ground state when $z_1$ is large. 
\end{sloppypar}

\subsection{Comparison with cavity-QED system}
In this section, we compare the stability and steady states of the coherent feedback network in Fig.~\ref{fig:scheme} and other coherent feedback network schemes  based on cavity-QED systems. For simplicity, we only consider the case where there is one two-level atom in resonance with the cavity in the coherent feedback network, which has been studied in \citep{photonfeedback,ding2022quantum,crowder2020quantum}. The cavity QED system can be represented with the Jaynes-Cummings model
\begin{equation} \label{con:HJC}
H_{\rm JC} = -g_c\left(a \sigma_1^+ + a^{\dag}\sigma_1^-\right), 
\end{equation}
where $a$ ($a^{\dag}$) is the annihilation(creation) operator for the cavity, and $g_c$ is the coupling strength between the atom and cavity. 
When the coherent feedback is realized by the cavity QED system coupled with a round-trip waveguide, the interaction Hamiltonian can be generalized from Eq.~(\ref{con:Hintj}) to \citep{ding2022quantum,photonfeedback,crowder2020quantum}
\begin{small}
\begin{equation} \label{con:HintjCav}
H_{\rm I}^{C-W} = H_{\rm JC} + \int  \left[g_{k1t}(k,t,z) d^{\dag}_k a  + g_{k1t}^*(k,t,z) d_k a^{\dag}\right]\mathrm{d}k,
\end{equation}
\end{small}%
with $d^{\dag}_k a$ and $d_k a^{\dag}$ representing the coupling or exchanging of photons between the waveguide and cavity.

\begin{sloppypar}
On the other hand, when the coherent feedback loop for the cavity QED system is closed by another cavity as~\citep{lang1973laser,gea1990treatment,gea2013space,photonfeedback} or Section  \uppercase\expandafter{\romannumeral3} of~\citep{ding2022quantum}, the interaction Hamiltonian for the coherent feedback network reads~\citep{photonfeedback,ding2022quantum}
\end{sloppypar}
\begin{small}
\begin{equation} \label{con:HintjCavDis}
H_{\rm I}^{C-C} = H_{\rm JC} + \sum_k \left[g_{k1t}\left(\frac{\omega_k}{c},t,z\right) d^{\dag}_k a  + g_{k1t}^*\left(\frac{\omega_k}{c},t,z\right) d_k a^{\dag}\right],
\end{equation}
\end{small}%
where the cavity for feedback can be modeled in analog with a waveguide with discrete modes~\citep{lang1973laser,photonfeedback,ding2022quantum}. We compare the atomic dynamics evaluated by $c_{e1}(t)$ in Eq.~(\ref{con:Ce1Ce2}) as follows.
\newtheorem{lemma}{Lemma}
\begin{lemma} \label{LemmaDZTAC1}
\citep{ding2022quantum,photonfeedback} For the coherent feedback network in Eq.~(\ref{con:HintjCav}), $c_{e1}(t)$ is exponentially stable when $z_1 \neq n\pi /\omega_a \ll c$, and is oscillating when $z_1 = n\pi /\omega_a \ll c$ for some  integer $n$.
\end{lemma}
\begin{lemma} \label{LemmaDZTAC2}
\citep{ding2022quantum} For the coherent feedback network in Eq.~(\ref{con:HintjCavDis}), $c_{e1}(t)$ is always oscillating for arbitrary $z_1$.
\end{lemma}
In the following, we compare the coherent feedback dynamics based on waveguide QED and cavity QED systems.
\begin{corollary} \label{FeedbackCompare}
\begin{sloppypar}
The coherent feedback network by a round-trip waveguide with continuous modes can induce exponentially decaying atomic states or persistently excited steady state, while can induce exponentially decaying states or oscillating states for a Jaynes-Cummings model without persistently excited atomic state.
\end{sloppypar}
\end{corollary}
\begin{Proof} 
The first half of the statement can be proved by combining \textbf{Theorems \ref{onephoton}-\ref{Smalleststability}}, and the second half of the statement can be proved by combining \textbf{Lemma~\ref{LemmaDZTAC1}} and \textbf{Lemma~\ref{LemmaDZTAC2}}. 
\end{Proof}

\begin{sloppypar}
The above comparisons illustrate that the coherent feedback loop realized by a waveguide makes it possible for the generation of stable photonic states via the spontaneous emission of atoms, while the terminal atomic state can be different according to whether the atom is coupled with a cavity or not. 
\end{sloppypar}

\section{Conclusion} \label{Sec:conclusion}
\begin{sloppypar}
In this paper, we study the  coherent feedback control dynamics with time delays based on atom-waveguide interactions. When the atoms are initially excited, the number of generated photons in the waveguide is influenced by atoms' positions and their chiral coupling strengths with different directional propagating modes in the waveguide. When the parameters above are properly designed, two-photon, one-photon and zero-photon states can be generated in the waveguide. This quantum coherent feedback dynamics based on waveguide QED can be analyzed from the perspective of linear control system with time delays, and large delays can induce excited steady atomic states, which is different from the circumstance that the coherent feedback delays are small. Moreover, the modeling of atom-waveguide interactions in the spatial domain can provide an alternative approach to investigate the coherent feedback dynamics, and the examples of coherent feedback with one or two atoms can provide a comprehensive study by combining the spatial and frequency domain analysis.
\end{sloppypar}

\appendix
$$\textbf{Appendix}$$
\section{Derivations of Eqs.~(\ref{con:Ceedelay},\ref{con:Cegdelay},\ref{con:Cgedelay})} \label{Sec:ChiralOdecal}
\begin{sloppypar}
In this appendix, we introduce the method for derivations of Eqs.~(\ref{con:Ceedelay},\ref{con:Cegdelay},\ref{con:Cgedelay}) in the main text. For example, to derive Eq.~(\ref{con:Cegdelay}), the second component at the right-hand side of Eq.~(\ref{model2}), $c_{kk}(k,k_1,t)$ can be replaced with the integration of Eq.~(\ref{model4}). Noticing that
\begin{small}
\begin{equation} \label{con:cegkPart2P1Chiral}
\begin{aligned}
&~~~~ \int c_{egk}(u,k)  g_{k1t}(k_1,u,z_1) g_{k1t}^*(k_1,t,z_1) \mathrm{d}k_1\\
&= c_{egk}(u,k) e^{-i\omega_a(u-t)} \left[\gamma_{1R}^2 \delta(u-t) + \gamma_{1L}^2 \delta(u-t)\right.\\
&\left.~~~- \gamma_{1L} \gamma_{1R} \delta\left(u-t+\frac{2z_1}{c}\right)  -\gamma_{1R}\gamma_{1L} \delta\left(u-t-\frac{2z_1}{c}\right) \right],
\end{aligned}
\end{equation}
\end{small}
one of the components in $\dot{c}_{egk}(t,k)$ reads
\begin{small}
\begin{equation} \label{con:cegkPart2P1step2Chiral}
\begin{aligned}
&~~~~ \int_0^t \mathrm{d}u \int c_{egk}(u,k)  g_{k1t}(k_1,u,z_1) g_{k1t}^*(k_1,t,z_1) \mathrm{d}k_1\\
&=\frac{\gamma_{1R}^2  + \gamma_{1L}^2 }{2} c_{egk}(t,k) - \gamma_{1L} \gamma_{1R} c_{egk}\left(t-\frac{2z_1}{c},k\right) e^{i\omega_a \frac{2z_1}{c}},
\end{aligned}
\end{equation}
\end{small}%
which means that the temporal evolution of $c_{egk}(t,k)$ is  influenced not only by the chiral coupling strengths, but also by the round-trip delay between the first atom and the mirror.
\end{sloppypar}

The following component in $\dot{c}_{egk}(t,k)$ yields
\begin{small}
\begin{equation} \label{con:cegkPart2P2Chiral}
\begin{aligned}
&~~~~\int c_{egk}(u,k_1)  g_{k1t}(k,u,z_1) g_{k1t}^*(k_1,t,z_1) \mathrm{d}k_1\\
&=\int c_{egk}(u,k_1)   \left\{i\gamma_{1R} e^{i\left[\left(\omega-\omega_a\right)u - \omega z_1/c\right]} - i\gamma_{1L}e^{i\left[\left(\omega-\omega_a\right)u + \omega z_1/c\right]}\right\}\\
&~~~~\left\{-i\gamma_{1R} e^{-i\left[\left(\omega_1-\omega_a\right)t - \omega_1 z_1/c\right]} + i\gamma_{1L}e^{-i\left[\left(\omega_1-\omega_a\right)t + \omega_1 z_1/c\right]}\right\}\mathrm{d}k_1,\\
\end{aligned}
\end{equation}
\end{small}%
which equals zero according to the following lemma.

\begin{lemma} \label{lemma1}
For the finite amplitude $c_{egk}(u,k_1)$ which is a continuous function of  the time variable $u$, we have
\[
\int c_{egk}(u,k_1)  e^{-i[(\omega_1-\omega_a)t - \omega_1 z_1/c]}\mathrm{d}k_1 = 0.
\]
\end{lemma}

\begin{Proof} Notice that
\begin{small}
\begin{equation} \label{con:Lemmaproof}
\begin{aligned}
&~~~~\int c_{egk}(u,k_1)  e^{-i[(\omega_1-\omega_a)t - \omega_1 z_1/c]}\mathrm{d}k_1 \\
&=e^{i\omega_a t} \left[\delta\left(t -  \frac{z_1}{c}\right)  c_{egk}(u,k_1)  - \int \delta\left(t -  \frac{z_1}{c}\right) \frac{\mathrm{d} c_{egk}(u,k_1)}{\mathrm{d} k_1}\mathrm{d}k_1 \right].\\
\end{aligned}
\end{equation}
\end{small}%
\begin{sloppypar}
When $t \neq z_1/c $, $\delta(t -  z_1/c) = 0$, consequently 
$\int c_{egk}(u,k_1)  e^{-i[(\omega_1-\omega_a)t - \omega_1 z_1/c]}\mathrm{d}k_1 = 0$. Because $c_{egk}$ is continuous, $\int c_{egk}(u,k_1)  e^{-i[(\omega_1-\omega_a)t - \omega_1 z_1/c]}\mathrm{d}k_1 = 0$ for arbitrary $u$.\qed
\end{sloppypar}
\end{Proof}

Besides, the following integration with respect to $k_1$ in $\dot{c}_{egk}(t,k)$ yields
\begin{footnotesize}
\begin{equation} \label{con:cegkPart2P3Chiral}
\begin{aligned}
&~~~~ \int c_{gek}(u,k)  g_{k2t}(k_1,u,z_2) g_{k1t}^*(k_1,t,z_1) \mathrm{d}k_1\\
&= c_{gek}(u,k) e^{-i\omega_a(u-t)}\left[\gamma_{1R}\gamma_{2R} \delta\left(u-t - \frac{z_2-z_1}{c}\right) - \gamma_{1R}\gamma_{2L} \delta \left(u-t + \frac{z_1+z_2}{c}\right)\right.\\
&~~~~\left.-\gamma_{2R}\gamma_{1L}  \delta\left(u-t - \frac{z_2+z_1}{c}\right)  + \gamma_{1L}\gamma_{2L} \delta\left(u-t + \frac{z_2-z_1}{c}\right) \right],
\end{aligned}
\end{equation}
\end{footnotesize}%
rendering that
\begin{scriptsize}
\begin{equation} \label{con:cegkPart2P3inttChiral}
\begin{aligned}
&~~~~ \int_0^t  \int c_{gek}(u,k)  g_{k2t}(k_1,u,z_2) g_{k1t}^*(k_1,t,z_1) \mathrm{d}k_1 \mathrm{d}u\\
&=-\gamma_{1R}\gamma_{2L} c_{gek}\left(t-\frac{z_1+z_2}{c},k\right) e^{i\omega_a \frac{z_1+z_2}{c}}+ \gamma_{1L}\gamma_{2L} c_{gek}\left(t-\frac{z_2-z_1}{c},k\right) e^{i\omega_a \frac{z_2-z_1}{c}}.
\end{aligned}
\end{equation}
\end{scriptsize}%
\begin{sloppypar}
\end{sloppypar}

Similar to the proof of \textbf{Lemma}~\ref{lemma1},
\[
\int c_{gek}(u,k_1)  g_{k2t}(k,u,z_2) g_{k1t}^*(k_1,t,z_1) \mathrm{d}k_1=0.
\]
Based on the calculations above, the dynamics of $c_{egk}(t,k)$ can be represented as in Eq.~(\ref{con:Cegdelay}). Besides, the derivations of Eq.~(\ref{con:Ceedelay},\ref{con:Cgedelay}) are similar to that of Eq.~(\ref{con:Cegdelay}), thus are omitted due to page limitations.

\section{Derivation of the interaction Hamiltonian in the spatial domain} \label{Sec:PositionHam}
When an atom is  coupled to an infinite waveguide, the photonic wave packet can be divided into the right-propagating and left-propagating components.  Accordingly, the Hamiltonian of the waveguide mode, given by the second term on the right-hand side of Eq.~(\ref{con:H0}) in the main text, can be divided into two counter-propagating parts as~\citep{boundary}
\begin{footnotesize}
\begin{equation} \label{con:wgtwodirection}
\begin{aligned}
\int_0^{\infty}  \omega_k d^{\dag}_kd_k \mathrm{d}k= &\int_0^{\infty}  \omega_{k_L} d^{\dag}_{k_L}d_{k_L}\mathrm{d}k_L  + \int_0^{\infty}  \omega_{k_R} d^{\dag}_{k_R}d_{k_R}\mathrm{d}k_R,
\end{aligned}
\end{equation}
\end{footnotesize}%
where $d^{\dag}_{k_L}$ ($d_{k_L}$) represents the creation (annihilation) operator of the left-moving photonic  wave packet, and $d^{\dag}_{k_R}(d_{k_R})$ represents that of the right-moving photonic  wave packet. Moreover, based on the linearization of the waveguide mode $\omega_k$ around a central frequency $\omega_0$, the left- and right-propagating modes can be represented as~\citep{Shentheory1,cyril1997kondo}
\begin{small}
\begin{equation} \label{con:Leftpropagating}
\left\{
\begin{aligned}
\int_0^{\infty}  \omega_{k_L} d^{\dag}_{k_L}d_{k_L}  \mathrm{d}k_L \simeq \int_0^{\infty}  \left[\omega_0 - v_g \left(k_L-k_0\right)\right] d^{\dag}_{k_L}d_{k_L}\mathrm{d}k_L,  \\
\int_0^{\infty}  \omega_{k_R} d^{\dag}_{k_R}d_{k_R} \mathrm{d}k_R \simeq \int_0^{\infty} \left[\omega_0 + v_g \left(k_R-k_0\right)\right] d^{\dag}_{k_R}d_{k_R}\mathrm{d}k_R,  
\end{aligned}
\right.
\end{equation}
\end{small}%
where $v_g$ is the group velocity of the field, $\omega_{k_L} \simeq \omega_0 - v_g (k_L-k_0)$ and $\omega_{k_R} \simeq \omega_0 + v_g (k_R-k_0)$. The creation and annihilation operators $d_{k_L}^{\dag}$ and $d_{k_L} $  for the left-propagating modes can be represented in terms of  the spatial-domain operators as
\begin{equation} \label{con:dxtranslate}
\left\{
\begin{aligned}
&d_{k_L}^{\dag} = \int_{-\infty}^{\infty}  c_L^{\dag}(z)e^{ik_L z}\mathrm{d}z,\\
&d_{k_L} = \int_{-\infty}^{\infty}  c_L(z) e^{-ik_L z}\mathrm{d}z,
\end{aligned}
\right.
\end{equation}
where $c_L^{\dag}(z)$ and $c_L(z)$ are the creation and annihilation operators for the left-propagating field in  the waveguide at position $z$, respectively.
Similarly, for the right-propagating fields,
\begin{equation} \label{con:dxtranslateRight}
\left\{
\begin{aligned}
&d_{k_R}^{\dag} = \int_{-\infty}^{\infty} c_R^{\dag}(z)e^{ik_R z} \mathrm{d}z,\\
&d_{k_R} = \int_{-\infty}^{\infty}  c_R(z)e^{-ik_R z}\mathrm{d}z, 
\end{aligned}
\right.
\end{equation}
where $c_R^{\dag}(z)$ and $c_R(z)$ are the creation and annihilation operators for the right-propagating field in the waveguide at position $z$.

\begin{sloppypar}
Based on the Hamiltonian in Eq.~(\ref{con:Hintj}) and the transformation between the frequency domain  and the spatial domain in Eqs.~(\ref{con:dxtranslate},\ref{con:dxtranslateRight}), the interaction Hamiltonian in the spatial domain is ~\citep{Shentheory1,boundary}
\begin{equation} \label{con:HIposition}
\begin{aligned}
&H_{\rm I} = -i  \int_{-\infty}^{\infty}  \left [\gamma_{1R} c_R(z) \delta(z-z_1) +\gamma_{1L} c_L(z) \delta(z-z_1) \right ] \sigma_1^+ \mathrm{d}z + \mathrm{H.c.},
\end{aligned}
\end{equation}
which gives the format of $H_{\rm I}$ in Eq. \eqref{con:HamiltonainZdomMarkoviaChiral0}. More details on the spatial modeling can be found in \citep{Shentheory1}.
\end{sloppypar}

\subsection{The Hamiltonian of the mirror}
The function of the mirror in Fig.~\ref{fig:scheme} is to reflect a left-moving photonic wave packet to a right-moving wave packet. Consider the circumstance that there are no atoms coupled with the waveguide. With the boundary of the mirror at $z=0$, the Hamiltonian of the system in the absence of the atoms is
\begin{equation} \label{con:HamnoAtom}
\begin{aligned}
&H'  =H_w + H_m,
\end{aligned}
\end{equation}
where $H_m$ represents the Hamiltonian of the mirror and $H_w$ is the waveguide Hamiltonian given in Eq. \eqref{con:HamWavePos}. Let
\begin{equation} \label{con:stateboundary}
\begin{aligned}
&|\Psi\rangle = \Theta(z)e^{ikz}c_R^{\dag}(z) |0\rangle + \Theta(z)e^{-ikz}c_L^{\dag}(z) |0\rangle,
\end{aligned}
\end{equation}
where $\Theta$ denotes the Heaviside step function.

\begin{definition}
\citep{boundary}~The quantum state $|\Psi\rangle$ in Eq. \eqref{con:stateboundary} is called an eigenstate corresponding to the eigen-spectrum $v_g k$ of the Hamiltonian $H'$   if $H'|\Psi\rangle = v_g k |\Psi\rangle$.
\end{definition}

\begin{lemma} \label{lemmaMirror}
The Hamiltonian of the mirror located at the left terminal $z=0$ is
\begin{small}
\begin{equation} \label{con:HamiltonainMirror1}
H_m =i 2v_g\int_{0^-}^{\infty}\left [ c_R^{\dag}(z)c_L(z)e^{2ikz} - c_L^{\dag}(z) c_R(z)e^{-2ikz} \right ]\delta(z) \mathrm{d}z.
\end{equation}
\end{small}
\end{lemma}

\begin{Proof}
The proof is constructive. For an eigenstate $|\Psi\rangle$ of the Hamiltonian $H' = H_m + H_w$, we have $(H_m + H_w)|\Psi\rangle = v_g k |\Psi\rangle $. Consequently, 
\begin{small}
\begin{equation} \label{con:Hmcalculate}
\begin{aligned}
&~~~~(v_g k -H_w) |\Psi\rangle\\
&=\bigg( v_g k -i v_g \int_{0}^{\infty} \mathrm{d}z c_L^{\dag}(z) \frac{\partial}{\partial z} c_L(z) \\
&+ i v_g\int_{0}^{\infty} \mathrm{d}z c_R^{\dag}(z) \frac{\partial}{\partial z} c_R(z)\bigg) [\Theta(z)e^{ikz}c_R^{\dag}(z) |0\rangle + \Theta(z)e^{-ikz}c_L^{\dag}(z) |0\rangle] \\
&= v_g k  |\Psi\rangle -kv_g |\Psi\rangle  -iv_g \delta(z)e^{-ikz}c_L^{\dag}(z)|0\rangle   + iv_g \delta(z)e^{ikz} c_R^{\dag}(z) |0\rangle\\
&= -iv_g \delta(z)e^{-ikz}c_L^{\dag}(z)|0\rangle  + iv_g \delta(z)e^{ikz} c_R^{\dag}(z) |0\rangle.\\
\end{aligned}
\end{equation}
\end{small}%
On the other hand, given the format of $H_m$ in Eq. \eqref{con:HamiltonainMirror1}, 
 we have
\begin{small}
\begin{equation} \label{con:CheckHm}
\begin{aligned}
&~~~~H_m|\Psi\rangle\\
&=\int_{0^-}^{\infty}\left[i 2v_g c_R^{\dag}(z)c_L(z)e^{2ikz}\delta(z) -i 2v_g c_L^{\dag}(z) c_R(z)e^{-2ikz}\delta(z) \right]\\
&~~~~\mathrm{d}z \left[\Theta(z)e^{ikz}c_R^{\dag}(z) |0\rangle + \Theta(z)e^{-ikz}c_L^{\dag}(z) |0\rangle\right]\\
&= i 2v_g \int_{0^-}^{\infty} c_R^{\dag}(z') e^{2ikz'}\delta(z') \delta(z-z') \mathrm{d}z' \Theta(z)e^{-ikz} |0\rangle \\
&~~~~- i 2v_g  \int_{0^-}^{\infty} c_L^{\dag}(z')e^{-2ikz}\delta(z')\delta(z-z')\mathrm{d}z' \Theta(z)e^{ikz} |0\rangle\\
&= i v_g  c_R^{\dag}(z) \delta(z)  \Theta(z)e^{ikz} |0\rangle - i v_g  c_L^{\dag}(z)\delta(z) \Theta(z)e^{-ikz} |0\rangle,\\
\end{aligned}
\end{equation}
\end{small}%
due to the fact that $\int_{0^-}^{\infty} \delta(z)\Theta(z) \mathrm{d}z = 1/2$. As a result, when $H_m$ is that in Eq. \eqref{con:Hmcalculate},  $H'|\Psi\rangle = v_g k |\Psi\rangle$ holds.
\qed
\end{Proof}

\begin{remark}
The mirror Hamiltonian is determined by the relative position between the mirror and waveguide. When the mirror is at the right terminal of the waveguide as adopted by \citep{boundary}, the function of the mirror is to reflect the right-propagating fields in the waveguide to the left-propagating fields, and the mirror Hamiltonian is different from the circumstance that the mirror is at the left terminal of the waveguide as in Eq.~(\ref{con:HamiltonainMirror1}).
\end{remark}

\section{One-atom model in the spatial domain}\label{Sec:oneatomApendix}
\begin{sloppypar}
The quantum state with one right- or left-propagating photon in the waveguide can be equivalently represented as in Eq. \eqref{con:statepacketInteraction} and Eq. \eqref{con:PositionfunctionStep} in the main text,  respectively. Substituting Eq.~(\ref{con:PositionfunctionStep_a}) into Eq.~(\ref{Posmodel2}) yields
\begin{footnotesize}
\begin{equation} \label{con:Takein1}
\begin{aligned}
&~~~~[\Theta(z) -\Theta(z-z_1)] \frac{\partial f_r(t-z/v_g)}{\partial t} +  \Theta(z-z_1) \frac{\partial g_r(t-z/v_g)}{\partial t} \\
& = -v_g [\delta(z) -\delta(z-z_1)] f_r\left(t-\frac{z}{v_g}\right)  -v_g [\Theta(z) -\Theta(z-z_1)] \frac{\partial f_r(t-z/v_g)}{\partial z} \\
&~~~ -v_g\delta(z-z_1) g_r\left(t-\frac{z}{v_g}\right) - v_g \Theta(z-z_1) \frac{\partial g_r(t-z/v_g)}{\partial z} \\
&~~~ + 2v_g\delta(z)  [\Theta(z) -\Theta(z-z_1)] f_l(t+z/v_g)  e^{2ikz} +\gamma_{1R}\delta(z-z_1)c_e(t)e^{-i\omega_a t}.\\
\end{aligned}
\end{equation}
\end{footnotesize}%
Noticing that $\frac{\partial f_r(t-z/v_g)}{\partial t} = -v_g \frac{\partial f_r(t-z/v_g)}{\partial z}$, and $\frac{\partial g_r(t-z/v_g)}{\partial t} = -v_g \frac{\partial g_r(t-z/v_g)}{\partial z}$, Eq.~(\ref{con:Takein1}) reads
\begin{footnotesize}
\begin{equation} \label{con:Takein12}
\begin{aligned}
&~~~~ [\delta(z) -\delta(z-z_1)] f_r\left(t-\frac{z}{v_g}\right) + \delta(z-z_1) g_r\left(t-\frac{z}{v_g}\right)  \\
& = 2\delta(z)  [\Theta(z) -\Theta(z-z_1)] f_l\left(t+\frac{z}{v_g}\right)e^{2ikz}  +\frac{\gamma_{1R}}{v_g}\delta(z-z_1)c_e(t)e^{-i\omega_a t}.\\
\end{aligned}
\end{equation}
\end{footnotesize}%
Integrate both sides of Eq. \eqref{con:Takein12} within $[z_1^-,z_1^+]$, we can derive that
\begin{small}
\begin{equation} \label{con:Takein12z1}
\begin{aligned}
g_r\left(t-\frac{z_1}{v_g}\right) -f_r\left(t-\frac{z_1}{v_g}\right)  = \frac{\gamma_{1R}}{v_g}c_e(t)e^{-i\omega_a t}.
\end{aligned}
\end{equation}
\end{small}%
\end{sloppypar}

\begin{sloppypar}
For the left-propagating mode, substituting Eq.~(\ref{con:PositionfunctionStep_b}) into Eq.~(\ref{Posmodel3}) yields
\begin{small}
\begin{equation} \label{con:Takein1Left}
\begin{aligned}
&~~~~\left [\Theta(z) -\Theta(z-z_1) \right ] \frac{\partial f_l(t+z/v_g)}{\partial t} \\
&=v_g\left [\delta(z) -\delta(z-z_1)\right ] f_l(t+z/v_g)  + v_g \left [\Theta(z) -\Theta(z-z_1) \right ] \frac{\partial f_l(t+z/v_g)}{\partial z}\\
 &~~~~+ \gamma_{1L}\delta(z-z_1)c_e(t)e^{-i\omega_a t} -2v_g\delta(z)  \Phi_R(z,t)e^{-2ikz},
\end{aligned}
\end{equation}
\end{small}%
where $\frac{\partial f_l(t+z/v_g)}{\partial t} = v_g \frac{\partial f_l(t+z/v_g)}{\partial z}$, and hence Eq.~(\ref{con:Takein1Left}) reads
\begin{small}
\begin{equation} \label{con:Takein1LeftControlEquation}
\begin{aligned}
&~~~~2\delta(z)  \Phi_R(z,t)e^{-2ikz} \\
&=[\delta(z) -\delta(z-z_1)] f_l\left(t+\frac{z}{v_g}\right) + \frac{\gamma_{1L}}{v_g}\delta(z-z_1)c_e(t)e^{-i\omega_a t}.\\
\end{aligned}
\end{equation}
\end{small}%
\end{sloppypar}

By integrating of Eq.~(\ref{con:Takein1LeftControlEquation}) over $\left[0^-,0^+\right]$  and $\left[ z_1^-, z_1^+ \right]$ respectively, we have
\begin{equation} \label{con:Leftwaveequation}
\left\{
\begin{aligned}
&v_g f_l\left(t+\frac{z_1}{v_g}\right) = \gamma_{1L} c_e(t)e^{-i\omega_a t},\\
&f_l(t) = 2 \Phi_R(0,t),
\end{aligned}
\right.
\end{equation}
and
\begin{equation} \label{con:flt}
\begin{aligned}
f_l(t) = \frac{\gamma_{1L}}{v_g} c_e(t-z_1/v_g)e^{-i\omega_a (t-z_1/v_g)}.
\end{aligned}
\end{equation}

Combine Eqs. \eqref{con:Takein12z1}, \eqref{con:flt} and \eqref{con:PositionfunctionStep}  with the boundary condition $\Phi_R(0,t) =-\Phi_L(0,t) $ at $z=0$, then $f_l(t) = -f_r(t)$ and
\begin{small}
\begin{equation} \label{con:grsolution}
\begin{aligned}
g_r(t) = &\frac{\gamma_{1R}}{v_g}c_e\left(t+\frac{z_1}{v_g}\right)e^{-i\omega_a \left(t+\frac{z_1}{v_g}\right)} - \frac{\gamma_{1L}}{v_g} c_e\left(t-\frac{z_1}{v_g}\right)e^{-i\omega_a \left(t-\frac{z_1}{v_g}\right)}.
\end{aligned}
\end{equation}
\end{small}%
Considering that the amplitudes for the right- and left-propagating photon wave packet  at $z=z_1$ can be represented as
\begin{subequations}  \label{con:PositionZ1Takein}
\begin{align}
&\Phi_R(z_1,t) =\frac{1}{2}f_r(t-z_1/v_g) + \frac{1}{2}g_r(t-z_1/v_g),\\
&\Phi_L(z_1,t) =\frac{1}{2}f_l(t+z_1/v_g),
\end{align}
\end{subequations}
with $\Theta(0) = 1/2$, substitute $g_r(t)$, $f_l(t)$ and $f_r(t)$ obtained above into Eq.~(\ref{Posmodel1}), we get
\begin{small}
\begin{equation} \label{con:CedelayCal}
\begin{aligned}
&~~~~\dot{c}_e(t) = -\left [\gamma_{1R} \Phi_R(z_1,t)  + \gamma_{1L}\Phi_L(z_1,t) \right ]e^{i\omega_a t}\\
&=-\frac{1}{2}\left [\gamma_{1R} \left(f_r\left(t-\frac{z_1}{v_g}\right) + g_r\left(t-\frac{z_1}{v_g}\right)\right) +  \gamma_{1L}f_l\left(t+\frac{z_1}{v_g}\right) \right ]e^{i\omega_a t},
\end{aligned}
\end{equation}
\end{small}%
resulting in Eq.~(\ref{con:MaintexCedelayCal}) in the main text.

\section{Two-atom model with one excitation in the spatial domain}\label{Sec:TwoatomOneExApendix}

Substituting the state representation in Eq. \eqref{con:TwoatomPositionfunctionStep} into Eq. \eqref{con:TwoatomOneEx} in the main text, we have
\begin{small}
\begin{subequations} \label{con:TwoatomDelta}
\begin{numcases}{}
\left[\delta(z) -\delta(z-z_1)\right] f_r\left(t-\frac{z}{v_g}\right)+   \delta(z-z_2) h_r\left(t-\frac{z}{v_g}\right)\notag\\
+ \left [ \delta(z-z_1) - \delta(z-z_2) \right ] g_r\left(t-\frac{z}{v_g}\right) \notag\\
= 2 \Phi_{g}^{l}(t,z) e^{2ikz} \delta(z) + \sum_{j=1,2} \frac{\gamma_{jR}}{v_g} c_j(t) \delta(z-z_j) e^{-i\omega_a t},\label{con:PositionTwoAOneE_aAPendix}\\
\left[\delta(z) -\delta(z-z_1)\right] f_l\left(t+\frac{z}{v_g}\right) +\left [ \delta(z-z_1) - \delta(z-z_2) \right ] g_l\left(t+\frac{z}{v_g}\right) \notag \\
= 2\delta(z)\Phi_{g}^{r}(t,z) e^{-2ikz} \delta(z)- \sum_{j=1,2} \frac{\gamma_{jL}}{v_g} c_j(t) \delta(z-z_j) e^{-i\omega_a t}.
\end{numcases}
\end{subequations}
\end{small}%
Consider the integration within $\left [ 0^- ,0^+\right ]$, $\left [ z_1^- ,z_1^+\right ]$ and $\left [ z_2^- ,z_2^+\right ]$ respectively, we have
\begin{small}
\begin{subequations}  \label{con:Intdelta}
\begin{align}
f_r(t) &=-f_l(t),\label{con:Eq2a1}\\
g_r\left(t-\frac{z_1}{v_g}\right) - f_r\left(t-\frac{z_1}{v_g}\right) &= \frac{\gamma_{1R}}{v_g} c_1(t)e^{-i\omega_at},\label{con:Eq2a2}\\
g_l\left(t+\frac{z_1}{v_g}\right) - f_l\left(t+\frac{z_1}{v_g}\right) &= -\frac{\gamma_{1L}}{v_g} c_1(t)e^{-i\omega_at},\label{con:Eq2a3}\\
h_r\left(t-\frac{z_2}{v_g}\right) -g_r\left(t-\frac{z_2}{v_g}\right) &= \frac{\gamma_{2R}}{v_g}c_2(t)e^{-i\omega_at},\label{con:Eq2a4}\\
g_l\left(t+\frac{z_2}{v_g}\right) &= \frac{\gamma_{2L}}{v_g} c_2(t)e^{-i\omega_at}. \label{con:Eq2a5}
\end{align}
\end{subequations}
\end{small}
Then similar to Eq.~(\ref{con:CedelayCal}),
\begin{small}
\begin{equation} \label{con:c1dotCal}
\begin{aligned}
&\dot{c}_1(t) =-\frac{\gamma_{1R}^2 + \gamma_{1L}^2}{2} c_1(t)  +\gamma_{1R}\gamma_{1L}c_1\left(t-\frac{2z_1}{v_g}\right)e^{i\omega_a\frac{2z_1}{v_g}} \\
&+\gamma_{1R}\gamma_{2L}c_2\left(t-\frac{z_1+z_2}{v_g}\right)e^{i\omega_a\frac{z_1+z_2}{v_g}}-\gamma_{1L}\gamma_{2L}c_2\left(t-\frac{z_2-z_1}{v_g}\right)e^{i\omega_a \frac{z_2-z_1}{v_g}},
\end{aligned}
\end{equation}
\end{small}%
and similarly $\dot{c}_2(t)$ can be derived as Eq.~(\ref{con:C2tOnephoton}).

Additionally, the photonic wave packet in the spatial domain can be solved based on  Eq.~(\ref{con:Intdelta}) as
\begin{small}
\begin{subequations} \label{con:Packetsolution}
\begin{align}
&g_l\left(t+\frac{z}{v_g}\right) = \frac{\gamma_{2L}}{v_g}c_2\left(t+\frac{z-z_2}{v_g}\right)e^{-i\omega_a\left(t+\frac{z-z_2}{v_g}\right)},\label{con:EdD7a}\\
&f_l\left(t+\frac{z}{v_g}\right) = \frac{\gamma_{2L}}{v_g}c_2\left(t+\frac{z-z_2}{v_g}\right)e^{-i\omega_a\left(t+\frac{z-z_2}{v_g}\right)} \notag\\
&~~~~~~~~~~~~~~~~~~~~~+ \frac{\gamma_{1L}}{v_g}c_1\left(t+\frac{z-z_1}{v_g}\right)e^{-i\omega_a\left(t+\frac{z-z_1}{v_g}\right)},\label{con:EdD7b}\\
&f_r\left(t-\frac{z}{v_g}\right) = - \frac{\gamma_{2L}}{v_g}c_2\left(t-\frac{z+z_2}{v_g}\right)e^{-i\omega_a\left(t-\frac{z+z_2}{v_g}\right)} \notag\\
&~~~~~~~~~~~~~~~~~~~~-\frac{\gamma_{1L}}{v_g}c_1\left(t-\frac{z+z_1}{v_g}\right)e^{-i\omega_a\left(t-\frac{z+z_1}{v_g}\right)},\label{con:EdD7c}\\
&g_r\left(t-\frac{z}{v_g}\right) = f_r\left(t-\frac{z}{v_g}\right)  +\frac{\gamma_{1R}}{v_g}c_1\left(t-\frac{z-z_1}{v_g}\right)e^{-i\omega_a\left(t-\frac{z-z_1}{v_g}\right)},\label{con:EdD7d}\\
&h_r\left(t-\frac{z}{v_g}\right) = g_r\left(t-\frac{z}{v_g}\right)  +\frac{\gamma_{2R}}{v_g}c_2\left(t-\frac{z-z_2}{v_g}\right)e^{-i\omega_a\left(t-\frac{z-z_2}{v_g}\right)}.\label{con:EdD7e}
\end{align}
\end{subequations}
\end{small}%

\section*{Acknowledgement}
\begin{sloppypar}
GFZ acknowledges supports from  Innovation Program for Quantum Science,  Technology 2023ZD0300600, Guangdong Provincial Quantum Science Strategic Initiative (No. GDZX2200001),  the Hong Kong Research Grant council
(RGC) Grants No. 15213924, and the CAS AMSS-PolyU Joint
Laboratory of Applied Mathematics. MTC and GQC acknowledge supports by National Nature Science Foundation of China with grant Nos.119075023, 12475010, and Key Project of Natural Science Foundation of Anhui Provincial Department of Education under Grant 2022AH040053.
\end{sloppypar}

\bibliographystyle{elsarticle-harv} 
\bibliography{waveguideTwoatom}

\begin{thebibliography}{67}
\expandafter\ifx\csname natexlab\endcsname\relax\def\natexlab#1{#1}\fi
\providecommand{\url}[1]{\texttt{#1}}
\providecommand{\href}[2]{#2}
\providecommand{\path}[1]{#1}
\providecommand{\DOIprefix}{doi:}
\providecommand{\ArXivprefix}{arXiv:}
\providecommand{\URLprefix}{URL: }
\providecommand{\Pubmedprefix}{pmid:}
\providecommand{\doi}[1]{\href{http://dx.doi.org/#1}{\path{#1}}}
\providecommand{\Pubmed}[1]{\href{pmid:#1}{\path{#1}}}
\providecommand{\bibinfo}[2]{#2}
\ifx\xfnm\relax \def\xfnm[#1]{\unskip,\space#1}\fi
\bibitem[{Almendros et~al.(2009)Almendros, Huwer, Piro, Rohde, Schuck,
  Hennrich, Dubin and Eschner}]{IonPRL}
\bibinfo{author}{Almendros, M.}, \bibinfo{author}{Huwer, J.},
  \bibinfo{author}{Piro, N.}, \bibinfo{author}{Rohde, F.},
  \bibinfo{author}{Schuck, C.}, \bibinfo{author}{Hennrich, M.},
  \bibinfo{author}{Dubin, F.}, \bibinfo{author}{Eschner, J.},
  \bibinfo{year}{2009}.
\newblock \bibinfo{title}{Bandwidth-tunable single-photon source in an ion-trap
  quantum network}.
\newblock \bibinfo{journal}{Physical Review Letters} \bibinfo{volume}{103},
  \bibinfo{pages}{213601}.
\bibitem[{Angulo et~al.(2019)Angulo, M{\'a}rquez and Bernal}]{angulo2019quasi}
\bibinfo{author}{Angulo, S.}, \bibinfo{author}{M{\'a}rquez, R.},
  \bibinfo{author}{Bernal, M.}, \bibinfo{year}{2019}.
\newblock \bibinfo{title}{Quasi-polynomial-based robust stability of time-delay
  systems can be less conservative than {Lyapunov--Krasovskii} approaches}.
\newblock \bibinfo{journal}{IEEE Transactions on Automatic Control}
  \bibinfo{volume}{65}, \bibinfo{pages}{3164--3169}.
\bibitem[{Barros et~al.(2009)Barros, Stute, Northup, Russo, Schmidt and
  Blatt}]{ion}
\bibinfo{author}{Barros, H.}, \bibinfo{author}{Stute, A.},
  \bibinfo{author}{Northup, T.}, \bibinfo{author}{Russo, C.},
  \bibinfo{author}{Schmidt, P.}, \bibinfo{author}{Blatt, R.},
  \bibinfo{year}{2009}.
\newblock \bibinfo{title}{Deterministic single-photon source from a single
  ion}.
\newblock \bibinfo{journal}{New Journal of Physics} \bibinfo{volume}{11},
  \bibinfo{pages}{103004}.
\bibitem[{Bradford and Shen(2013)}]{boundary}
\bibinfo{author}{Bradford, M.}, \bibinfo{author}{Shen, J.T.},
  \bibinfo{year}{2013}.
\newblock \bibinfo{title}{Spontaneous emission in cavity {QED} with a
  terminated waveguide}.
\newblock \bibinfo{journal}{Physical Review A} \bibinfo{volume}{87},
  \bibinfo{pages}{063830}.
\bibitem[{Bradford and Shen(2015)}]{bradford2015architecture}
\bibinfo{author}{Bradford, M.}, \bibinfo{author}{Shen, J.T.},
  \bibinfo{year}{2015}.
\newblock \bibinfo{title}{Architecture dependence of photon antibunching in
  cavity quantum electrodynamics}.
\newblock \bibinfo{journal}{Physical Review A} \bibinfo{volume}{92},
  \bibinfo{pages}{023810}.
\bibitem[{Cardona et~al.(2020)Cardona, Sarlette and
  Rouchon}]{Pierre2020exponential}
\bibinfo{author}{Cardona, G.}, \bibinfo{author}{Sarlette, A.},
  \bibinfo{author}{Rouchon, P.}, \bibinfo{year}{2020}.
\newblock \bibinfo{title}{Exponential stabilization of quantum systems under
  continuous non-demolition measurements}.
\newblock \bibinfo{journal}{Automatica} \bibinfo{volume}{112},
  \bibinfo{pages}{108719}.
\bibitem[{Chen et~al.(2017)Chen, Zhou and Shen}]{ShentwophtonP}
\bibinfo{author}{Chen, Z.}, \bibinfo{author}{Zhou, Y.}, \bibinfo{author}{Shen,
  J.T.}, \bibinfo{year}{2017}.
\newblock \bibinfo{title}{Dissipation-induced photonic-correlation transition
  in {waveguide-QED} systems}.
\newblock \bibinfo{journal}{Physical Review A} \bibinfo{volume}{96},
  \bibinfo{pages}{053805}.
\bibitem[{Cheng et~al.(2017)Cheng, Xu and Agarwal}]{cheng2017waveguide}
\bibinfo{author}{Cheng, M.T.}, \bibinfo{author}{Xu, J.},
  \bibinfo{author}{Agarwal, G.S.}, \bibinfo{year}{2017}.
\newblock \bibinfo{title}{Waveguide transport mediated by strong coupling with
  atoms}.
\newblock \bibinfo{journal}{Physical Review A} \bibinfo{volume}{95},
  \bibinfo{pages}{053807}.
\bibitem[{Crowder et~al.(2020)Crowder, Carmichael and
  Hughes}]{crowder2020quantum}
\bibinfo{author}{Crowder, G.}, \bibinfo{author}{Carmichael, H.},
  \bibinfo{author}{Hughes, S.}, \bibinfo{year}{2020}.
\newblock \bibinfo{title}{Quantum trajectory theory of few-photon {cavity-QED}
  systems with a time-delayed coherent feedback}.
\newblock \bibinfo{journal}{Physical Review A} \bibinfo{volume}{101},
  \bibinfo{pages}{023807}.
\bibitem[{Cyril~Hewson(1997)}]{cyril1997kondo}
\bibinfo{author}{Cyril~Hewson, A.}, \bibinfo{year}{1997}.
\newblock \bibinfo{title}{The Kondo Problem to Heavy Fermions}.
\bibitem[{Dinc(2020)}]{dinc2020diagrammatic}
\bibinfo{author}{Dinc, F.}, \bibinfo{year}{2020}.
\newblock \bibinfo{title}{Diagrammatic approach for analytical {non-Markovian}
  time evolution: Fermi's two-atom problem and causality in waveguide quantum
  electrodynamics}.
\newblock \bibinfo{journal}{Physical Review A} \bibinfo{volume}{102},
  \bibinfo{pages}{013727}.
\bibitem[{Ding et~al.(2025)Ding, Amini, Zhang and Gough}]{ding2023quantum}
\bibinfo{author}{Ding, H.}, \bibinfo{author}{Amini, N.H.},
  \bibinfo{author}{Zhang, G.}, \bibinfo{author}{Gough, J.E.},
  \bibinfo{year}{2025}.
\newblock \bibinfo{title}{Quantum coherent and measurement feedback control
  based on atoms coupled with a semi-infinite waveguide}.
\newblock \bibinfo{journal}{SIAM Journal on Control and Optimization}
  \bibinfo{volume}{63}, \bibinfo{pages}{S231--S257}.
\bibitem[{Ding and Zhang(2023)}]{ding2022quantum}
\bibinfo{author}{Ding, H.}, \bibinfo{author}{Zhang, G.}, \bibinfo{year}{2023}.
\newblock \bibinfo{title}{Quantum coherent feedback control with photons}.
\newblock \bibinfo{journal}{IEEE Transactions on Automatic Control}
  \bibinfo{volume}{69}, \bibinfo{pages}{856--871}.
\bibitem[{Domokos et~al.(2002)Domokos, Horak and Ritsch}]{AtomWaveguideHam}
\bibinfo{author}{Domokos, P.}, \bibinfo{author}{Horak, P.},
  \bibinfo{author}{Ritsch, H.}, \bibinfo{year}{2002}.
\newblock \bibinfo{title}{Quantum description of light-pulse scattering on a
  single atom in waveguides}.
\newblock \bibinfo{journal}{Physical Review A} \bibinfo{volume}{65},
  \bibinfo{pages}{033832}.
\bibitem[{Dorner and Zoller(2002)}]{halfcavity}
\bibinfo{author}{Dorner, U.}, \bibinfo{author}{Zoller, P.},
  \bibinfo{year}{2002}.
\newblock \bibinfo{title}{Laser-driven atoms in half-cavities}.
\newblock \bibinfo{journal}{Physical Review A} \bibinfo{volume}{66},
  \bibinfo{pages}{023816}.
\bibitem[{Flamini et~al.(2018)Flamini, Spagnolo and
  Sciarrino}]{flamini2018photonic}
\bibinfo{author}{Flamini, F.}, \bibinfo{author}{Spagnolo, N.},
  \bibinfo{author}{Sciarrino, F.}, \bibinfo{year}{2018}.
\newblock \bibinfo{title}{Photonic quantum information processing: a review}.
\newblock \bibinfo{journal}{Reports on Progress in Physics}
  \bibinfo{volume}{82}, \bibinfo{pages}{016001}.
\bibitem[{Gea-Banacloche(2013)}]{gea2013space}
\bibinfo{author}{Gea-Banacloche, J.}, \bibinfo{year}{2013}.
\newblock \bibinfo{title}{Space-time descriptions of quantum fields interacting
  with optical cavities}.
\newblock \bibinfo{journal}{Physical Review A} \bibinfo{volume}{87},
  \bibinfo{pages}{023832}.
\bibitem[{Gea-Banacloche et~al.(1990)Gea-Banacloche, Lu, Pedrotti, Prasad,
  Scully and W{\'o}dkiewicz}]{gea1990treatment}
\bibinfo{author}{Gea-Banacloche, J.}, \bibinfo{author}{Lu, N.},
  \bibinfo{author}{Pedrotti, L.M.}, \bibinfo{author}{Prasad, S.},
  \bibinfo{author}{Scully, M.O.}, \bibinfo{author}{W{\'o}dkiewicz, K.},
  \bibinfo{year}{1990}.
\newblock \bibinfo{title}{Treatment of the spectrum of squeezing based on the
  modes of the universe. {I. Theory} and a physical picture}.
\newblock \bibinfo{journal}{Physical Review A} \bibinfo{volume}{41},
  \bibinfo{pages}{369}.
\bibitem[{Gonzalez-Ballestero et~al.(2015)Gonzalez-Ballestero, Gonzalez-Tudela,
  Garcia-Vidal and Moreno}]{chiralRoute}
\bibinfo{author}{Gonzalez-Ballestero, C.}, \bibinfo{author}{Gonzalez-Tudela,
  A.}, \bibinfo{author}{Garcia-Vidal, F.J.}, \bibinfo{author}{Moreno, E.},
  \bibinfo{year}{2015}.
\newblock \bibinfo{title}{Chiral route to spontaneous entanglement generation}.
\newblock \bibinfo{journal}{Physical Review B} \bibinfo{volume}{92},
  \bibinfo{pages}{155304}.
\bibitem[{Guimond et~al.(2017)Guimond, Pletyukhov, Pichler and
  Zoller}]{guimond2017delayed}
\bibinfo{author}{Guimond, P.O.}, \bibinfo{author}{Pletyukhov, M.},
  \bibinfo{author}{Pichler, H.}, \bibinfo{author}{Zoller, P.},
  \bibinfo{year}{2017}.
\newblock \bibinfo{title}{Delayed coherent quantum feedback from a scattering
  theory and a matrix product state perspective}.
\newblock \bibinfo{journal}{Quantum Science and Technology}
  \bibinfo{volume}{2}, \bibinfo{pages}{044012}.
\bibitem[{Hijlkema et~al.(2007)Hijlkema, Weber, Specht, Webster, Kuhn and
  Rempe}]{neutralatom}
\bibinfo{author}{Hijlkema, M.}, \bibinfo{author}{Weber, B.},
  \bibinfo{author}{Specht, H.P.}, \bibinfo{author}{Webster, S.C.},
  \bibinfo{author}{Kuhn, A.}, \bibinfo{author}{Rempe, G.},
  \bibinfo{year}{2007}.
\newblock \bibinfo{title}{A single-photon server with just one atom}.
\newblock \bibinfo{journal}{Nature Physics} \bibinfo{volume}{3},
  \bibinfo{pages}{253--255}.
\bibitem[{Houck et~al.(2007)Houck, Schuster, Gambetta, Schreier, Johnson, Chow,
  Frunzio, Majer, Devoret, Girvin et~al.}]{fly}
\bibinfo{author}{Houck, A.A.}, \bibinfo{author}{Schuster, D.},
  \bibinfo{author}{Gambetta, J.}, \bibinfo{author}{Schreier, J.},
  \bibinfo{author}{Johnson, B.}, \bibinfo{author}{Chow, J.},
  \bibinfo{author}{Frunzio, L.}, \bibinfo{author}{Majer, J.},
  \bibinfo{author}{Devoret, M.}, \bibinfo{author}{Girvin, S.}, et~al.,
  \bibinfo{year}{2007}.
\newblock \bibinfo{title}{Generating single microwave photons in a circuit}.
\newblock \bibinfo{journal}{Nature} \bibinfo{volume}{449},
  \bibinfo{pages}{328--331}.
\bibitem[{Hu et~al.(2018)Hu, Zou and Zhang}]{InfiniteWaveguidePFeq}
\bibinfo{author}{Hu, Q.}, \bibinfo{author}{Zou, B.}, \bibinfo{author}{Zhang,
  Y.}, \bibinfo{year}{2018}.
\newblock \bibinfo{title}{Transmission and correlation of a two-photon pulse in
  a one-dimensional waveguide coupled with quantum emitters}.
\newblock \bibinfo{journal}{Physical Review A} \bibinfo{volume}{97},
  \bibinfo{pages}{033847}.
\bibitem[{Huo and Li(2020)}]{huo2020absorption}
\bibinfo{author}{Huo, M.}, \bibinfo{author}{Li, Y.}, \bibinfo{year}{2020}.
\newblock \bibinfo{title}{Absorption and delayed reemission in an array of
  atoms strongly coupled to a waveguide}.
\newblock \bibinfo{journal}{Physical Review A} \bibinfo{volume}{102},
  \bibinfo{pages}{033728}.
\bibitem[{Kamen(1980)}]{kamen1980relationship}
\bibinfo{author}{Kamen, E.}, \bibinfo{year}{1980}.
\newblock \bibinfo{title}{On the relationship between zero criteria for
  two-variable polynomials and asymptotic stability of delay differential
  equations}.
\newblock \bibinfo{journal}{IEEE Transactions on Automatic Control}
  \bibinfo{volume}{25}, \bibinfo{pages}{983--984}.
\bibitem[{Kamen(1982)}]{kamen1982linear}
\bibinfo{author}{Kamen, E.}, \bibinfo{year}{1982}.
\newblock \bibinfo{title}{Linear systems with commensurate time delays:
  Stability and stabilization independent of delay}.
\newblock \bibinfo{journal}{IEEE Transactions on Automatic Control}
  \bibinfo{volume}{27}, \bibinfo{pages}{367--375}.
\bibitem[{Kashima and Yamamoto(2009)}]{YamamotoFeedbackDelay}
\bibinfo{author}{Kashima, K.}, \bibinfo{author}{Yamamoto, N.},
  \bibinfo{year}{2009}.
\newblock \bibinfo{title}{Control of quantum systems despite feedback delay}.
\newblock \bibinfo{journal}{IEEE Transactions on Automatic Control}
  \bibinfo{volume}{54}, \bibinfo{pages}{876--881}.
\bibitem[{Keller et~al.(2004)Keller, Lange, Hayasaka, Lange and
  Walther}]{keller2004continuous}
\bibinfo{author}{Keller, M.}, \bibinfo{author}{Lange, B.},
  \bibinfo{author}{Hayasaka, K.}, \bibinfo{author}{Lange, W.},
  \bibinfo{author}{Walther, H.}, \bibinfo{year}{2004}.
\newblock \bibinfo{title}{Continuous generation of single photons with
  controlled waveform in an ion-trap cavity system}.
\newblock \bibinfo{journal}{Nature} \bibinfo{volume}{431},
  \bibinfo{pages}{1075--1078}.
\bibitem[{Kharitonov and Zhabko(2002)}]{kharitonov2002robust}
\bibinfo{author}{Kharitonov, V.L.}, \bibinfo{author}{Zhabko, A.P.},
  \bibinfo{year}{2002}.
\newblock \bibinfo{title}{Robust stability of time-delay systems}.
\newblock \bibinfo{journal}{IEEE Transactions on Automatic Control}
  \bibinfo{volume}{39}, \bibinfo{pages}{2388--2397}.
\bibitem[{Kockum et~al.(2018)Kockum, Johansson and Nori}]{Anton2018PRL}
\bibinfo{author}{Kockum, A.F.}, \bibinfo{author}{Johansson, G.},
  \bibinfo{author}{Nori, F.}, \bibinfo{year}{2018}.
\newblock \bibinfo{title}{Decoherence-free interaction between giant atoms in
  waveguide quantum electrodynamics}.
\newblock \bibinfo{journal}{Physical Review Letters} \bibinfo{volume}{120},
  \bibinfo{pages}{140404}.
\bibitem[{Kuhn et~al.(2002)Kuhn, Hennrich and Rempe}]{threeLatom}
\bibinfo{author}{Kuhn, A.}, \bibinfo{author}{Hennrich, M.},
  \bibinfo{author}{Rempe, G.}, \bibinfo{year}{2002}.
\newblock \bibinfo{title}{Deterministic single-photon source for distributed
  quantum networking}.
\newblock \bibinfo{journal}{Physical Review Letters} \bibinfo{volume}{89},
  \bibinfo{pages}{067901}.
\bibitem[{La and Ranjan(2007)}]{la2007asymptotic}
\bibinfo{author}{La, R.J.}, \bibinfo{author}{Ranjan, P.}, \bibinfo{year}{2007}.
\newblock \bibinfo{title}{Asymptotic stability of a rate control system with
  communication delays}.
\newblock \bibinfo{journal}{IEEE Transactions on Automatic Control}
  \bibinfo{volume}{52}, \bibinfo{pages}{1920--1925}.
\bibitem[{Lang et~al.(1973)Lang, Scully and Lamb~Jr}]{lang1973laser}
\bibinfo{author}{Lang, R.}, \bibinfo{author}{Scully, M.O.},
  \bibinfo{author}{Lamb~Jr, W.E.}, \bibinfo{year}{1973}.
\newblock \bibinfo{title}{Why is the laser line so narrow? {A} theory of
  single-quasimode laser operation}.
\newblock \bibinfo{journal}{Physical Review A} \bibinfo{volume}{7},
  \bibinfo{pages}{1788}.
\bibitem[{Li and Song(2016)}]{li2016stabilization}
\bibinfo{author}{Li, X.}, \bibinfo{author}{Song, S.}, \bibinfo{year}{2016}.
\newblock \bibinfo{title}{Stabilization of delay systems: delay-dependent
  impulsive control}.
\newblock \bibinfo{journal}{IEEE Transactions on Automatic Control}
  \bibinfo{volume}{62}, \bibinfo{pages}{406--411}.
\bibitem[{Lodahl et~al.(2017)Lodahl, Mahmoodian, Stobbe, Rauschenbeutel,
  Schneeweiss, Volz, Pichler and Zoller}]{lodahl2017chiral}
\bibinfo{author}{Lodahl, P.}, \bibinfo{author}{Mahmoodian, S.},
  \bibinfo{author}{Stobbe, S.}, \bibinfo{author}{Rauschenbeutel, A.},
  \bibinfo{author}{Schneeweiss, P.}, \bibinfo{author}{Volz, J.},
  \bibinfo{author}{Pichler, H.}, \bibinfo{author}{Zoller, P.},
  \bibinfo{year}{2017}.
\newblock \bibinfo{title}{Chiral quantum optics}.
\newblock \bibinfo{journal}{Nature} \bibinfo{volume}{541},
  \bibinfo{pages}{473--480}.
\bibitem[{Malakhovski and Mirkin(2006)}]{malakhovski2006stability}
\bibinfo{author}{Malakhovski, E.}, \bibinfo{author}{Mirkin, L.},
  \bibinfo{year}{2006}.
\newblock \bibinfo{title}{On stability of second-order quasi-polynomials with a
  single delay}.
\newblock \bibinfo{journal}{Automatica} \bibinfo{volume}{42},
  \bibinfo{pages}{1041--1047}.
\bibitem[{Michler et~al.(2000)Michler, Kiraz, Becher, Schoenfeld, Petroff,
  Zhang, Hu and Imamoglu}]{michler2000quantum}
\bibinfo{author}{Michler, P.}, \bibinfo{author}{Kiraz, A.},
  \bibinfo{author}{Becher, C.}, \bibinfo{author}{Schoenfeld, W.},
  \bibinfo{author}{Petroff, P.}, \bibinfo{author}{Zhang, L.},
  \bibinfo{author}{Hu, E.}, \bibinfo{author}{Imamoglu, A.},
  \bibinfo{year}{2000}.
\newblock \bibinfo{title}{A quantum dot single-photon turnstile device}.
\newblock \bibinfo{journal}{Science} \bibinfo{volume}{290},
  \bibinfo{pages}{2282--2285}.
\bibitem[{Mirza and Schotland(2016)}]{WaveguideNomirror}
\bibinfo{author}{Mirza, I.M.}, \bibinfo{author}{Schotland, J.C.},
  \bibinfo{year}{2016}.
\newblock \bibinfo{title}{Two-photon entanglement in multiqubit
  bidirectional-waveguide {QED}}.
\newblock \bibinfo{journal}{Physical Review A} \bibinfo{volume}{94},
  \bibinfo{pages}{012309}.
\bibitem[{Monroe(2002)}]{monroe2002quantum}
\bibinfo{author}{Monroe, C.}, \bibinfo{year}{2002}.
\newblock \bibinfo{title}{Quantum information processing with atoms and
  photons}.
\newblock \bibinfo{journal}{Nature} \bibinfo{volume}{416},
  \bibinfo{pages}{238--246}.
\bibitem[{Moradian and Kia(2019)}]{moradian2019positive}
\bibinfo{author}{Moradian, H.}, \bibinfo{author}{Kia, S.S.},
  \bibinfo{year}{2019}.
\newblock \bibinfo{title}{On the positive effect of delay on the rate of
  convergence of a class of linear time-delayed systems}.
\newblock \bibinfo{journal}{IEEE Transactions on Automatic Control}
  \bibinfo{volume}{65}, \bibinfo{pages}{4832--4839}.
\bibitem[{N\'emet et~al.(2019)N\'emet, Carmele, Parkins and
  Knorr}]{photonfeedback}
\bibinfo{author}{N\'emet, N.}, \bibinfo{author}{Carmele, A.},
  \bibinfo{author}{Parkins, S.}, \bibinfo{author}{Knorr, A.},
  \bibinfo{year}{2019}.
\newblock \bibinfo{title}{Comparison between continuous- and discrete-mode
  coherent feedback for the {Jaynes-Cummings} model}.
\newblock \bibinfo{journal}{Physical Review A} \bibinfo{volume}{100},
  \bibinfo{pages}{023805}.
\bibitem[{Northup and Blatt(2014)}]{northup2014quantum}
\bibinfo{author}{Northup, T.}, \bibinfo{author}{Blatt, R.},
  \bibinfo{year}{2014}.
\newblock \bibinfo{title}{Quantum information transfer using photons}.
\newblock \bibinfo{journal}{Nature Photonics} \bibinfo{volume}{8},
  \bibinfo{pages}{356--363}.
\bibitem[{Olbrot(1984)}]{TAC1984olbrot}
\bibinfo{author}{Olbrot, A.}, \bibinfo{year}{1984}.
\newblock \bibinfo{title}{A sufficiently large time delay in feedback loop must
  destroy exponential stability of any decay rate}.
\newblock \bibinfo{journal}{IEEE Transactions on Automatic Control}
  \bibinfo{volume}{29}, \bibinfo{pages}{367--368}.
\bibitem[{Peng et~al.(2016)Peng, De~Graaf, Tsai and
  Astafiev}]{peng2016tuneable}
\bibinfo{author}{Peng, Z.}, \bibinfo{author}{De~Graaf, S.},
  \bibinfo{author}{Tsai, J.}, \bibinfo{author}{Astafiev, O.},
  \bibinfo{year}{2016}.
\newblock \bibinfo{title}{Tuneable on-demand single-photon source in the
  microwave range}.
\newblock \bibinfo{journal}{Nature Communications} \bibinfo{volume}{7},
  \bibinfo{pages}{1--6}.
\bibitem[{Pichler and Zoller(2016)}]{ZollerPRL}
\bibinfo{author}{Pichler, H.}, \bibinfo{author}{Zoller, P.},
  \bibinfo{year}{2016}.
\newblock \bibinfo{title}{Photonic circuits with time delays and quantum
  feedback}.
\newblock \bibinfo{journal}{Physical Review Letters} \bibinfo{volume}{116},
  \bibinfo{pages}{093601}.
\bibitem[{Regidor et~al.(2021)Regidor, Crowder, Carmichael and
  Hughes}]{WaveguideModelingCan}
\bibinfo{author}{Regidor, S.A.}, \bibinfo{author}{Crowder, G.},
  \bibinfo{author}{Carmichael, H.}, \bibinfo{author}{Hughes, S.},
  \bibinfo{year}{2021}.
\newblock \bibinfo{title}{Modeling quantum light-matter interactions in
  waveguide {QED} with retardation, nonlinear interactions, and a time-delayed
  feedback: Matrix product states versus a space-discretized waveguide model}.
\newblock \bibinfo{journal}{Physical Review Research} \bibinfo{volume}{3},
  \bibinfo{pages}{023030}.
\bibitem[{Shao(2009)}]{shao2009new}
\bibinfo{author}{Shao, H.}, \bibinfo{year}{2009}.
\newblock \bibinfo{title}{New delay-dependent stability criteria for systems
  with interval delay}.
\newblock \bibinfo{journal}{Automatica} \bibinfo{volume}{45},
  \bibinfo{pages}{744--749}.
\bibitem[{Shen and Fan(2009)}]{Shentheory1}
\bibinfo{author}{Shen, J.T.}, \bibinfo{author}{Fan, S.}, \bibinfo{year}{2009}.
\newblock \bibinfo{title}{Theory of single-photon transport in a single-mode
  waveguide. {I. Coupling} to a cavity containing a two-level atom}.
\newblock \bibinfo{journal}{Physical Review A} \bibinfo{volume}{79},
  \bibinfo{pages}{023837}.
\bibitem[{Simon(2017)}]{simon2017towards}
\bibinfo{author}{Simon, C.}, \bibinfo{year}{2017}.
\newblock \bibinfo{title}{Towards a global quantum network}.
\newblock \bibinfo{journal}{Nature Photonics} \bibinfo{volume}{11},
  \bibinfo{pages}{678--680}.
\bibitem[{Sinha et~al.(2020)Sinha, Gonz{\'a}lez-Tudela, Lu and
  Solano}]{sinha2020collective}
\bibinfo{author}{Sinha, K.}, \bibinfo{author}{Gonz{\'a}lez-Tudela, A.},
  \bibinfo{author}{Lu, Y.}, \bibinfo{author}{Solano, P.}, \bibinfo{year}{2020}.
\newblock \bibinfo{title}{Collective radiation from distant emitters}.
\newblock \bibinfo{journal}{Physical Review A} \bibinfo{volume}{102},
  \bibinfo{pages}{043718}.
\bibitem[{Soro and Kockum(2022)}]{soro2022chiral}
\bibinfo{author}{Soro, A.}, \bibinfo{author}{Kockum, A.F.},
  \bibinfo{year}{2022}.
\newblock \bibinfo{title}{Chiral quantum optics with giant atoms}.
\newblock \bibinfo{journal}{Physical Review A} \bibinfo{volume}{105},
  \bibinfo{pages}{023712}.
\bibitem[{Sun et~al.(2010)Sun, Liu, Chen and Rees}]{sun2010improved}
\bibinfo{author}{Sun, J.}, \bibinfo{author}{Liu, G.}, \bibinfo{author}{Chen,
  J.}, \bibinfo{author}{Rees, D.}, \bibinfo{year}{2010}.
\newblock \bibinfo{title}{Improved delay-range-dependent stability criteria for
  linear systems with time-varying delays}.
\newblock \bibinfo{journal}{Automatica} \bibinfo{volume}{46},
  \bibinfo{pages}{466--470}.
\bibitem[{Tan et~al.(2011)Tan, Zhang and Li}]{entanglecavity}
\bibinfo{author}{Tan, H.T.}, \bibinfo{author}{Zhang, W.M.},
  \bibinfo{author}{Li, G.x.}, \bibinfo{year}{2011}.
\newblock \bibinfo{title}{Entangling two distant nanocavities via a waveguide}.
\newblock \bibinfo{journal}{Physical Review A} \bibinfo{volume}{83},
  \bibinfo{pages}{062310}.
\bibitem[{Tao et~al.(2022)Tao, Xiao, Zheng, Zhou, Ding, Jiang and
  Wu}]{tao2022design}
\bibinfo{author}{Tao, B.}, \bibinfo{author}{Xiao, M.}, \bibinfo{author}{Zheng,
  W.X.}, \bibinfo{author}{Zhou, Y.}, \bibinfo{author}{Ding, J.},
  \bibinfo{author}{Jiang, G.}, \bibinfo{author}{Wu, X.}, \bibinfo{year}{2022}.
\newblock \bibinfo{title}{Design and dynamics analysis of a time-delay feedback
  controller with distributed characteristic}.
\newblock \bibinfo{journal}{IEEE Transactions on Automatic Control}
  \bibinfo{volume}{68}, \bibinfo{pages}{1926--1933}.
\bibitem[{Tufarelli et~al.(2013)Tufarelli, Ciccarello and
  Kim}]{waveguideOneatom}
\bibinfo{author}{Tufarelli, T.}, \bibinfo{author}{Ciccarello, F.},
  \bibinfo{author}{Kim, M.S.}, \bibinfo{year}{2013}.
\newblock \bibinfo{title}{Dynamics of spontaneous emission in a single-end
  photonic waveguide}.
\newblock \bibinfo{journal}{Physical Review A} \bibinfo{volume}{87},
  \bibinfo{pages}{013820}.
\bibitem[{Xu and Lam(2005)}]{xu2005improved}
\bibinfo{author}{Xu, S.}, \bibinfo{author}{Lam, J.}, \bibinfo{year}{2005}.
\newblock \bibinfo{title}{Improved delay-dependent stability criteria for
  time-delay systems}.
\newblock \bibinfo{journal}{IEEE Transactions on Automatic Control}
  \bibinfo{volume}{50}, \bibinfo{pages}{384--387}.
\bibitem[{Yamamoto(2014)}]{YamamotoPRX}
\bibinfo{author}{Yamamoto, N.}, \bibinfo{year}{2014}.
\newblock \bibinfo{title}{Coherent versus measurement feedback: Linear systems
  theory for quantum information}.
\newblock \bibinfo{journal}{Physical Review X} \bibinfo{volume}{4},
  \bibinfo{pages}{041029}.
\bibitem[{Yan et~al.(2011)Yan, Wei, Jia and Shen}]{yan2011controlling}
\bibinfo{author}{Yan, C.H.}, \bibinfo{author}{Wei, L.F.}, \bibinfo{author}{Jia,
  W.Z.}, \bibinfo{author}{Shen, J.T.}, \bibinfo{year}{2011}.
\newblock \bibinfo{title}{Controlling resonant photonic transport along optical
  waveguides by two-level atoms}.
\newblock \bibinfo{journal}{Physical Review A} \bibinfo{volume}{84},
  \bibinfo{pages}{045801}.
\bibitem[{Zhang et~al.(2020)Zhang, You and Lu}]{ZhangBin}
\bibinfo{author}{Zhang, B.}, \bibinfo{author}{You, S.}, \bibinfo{author}{Lu,
  M.}, \bibinfo{year}{2020}.
\newblock \bibinfo{title}{Enhancement of spontaneous entanglement generation
  via coherent quantum feedback}.
\newblock \bibinfo{journal}{Physical Review A} \bibinfo{volume}{101},
  \bibinfo{pages}{032335}.
\bibitem[{Zhang(2020)}]{zhang2020single}
\bibinfo{author}{Zhang, G.}, \bibinfo{year}{2020}.
\newblock \bibinfo{title}{Single-photon coherent feedback control and
  filtering}.
\newblock \bibinfo{journal}{Encyclopedia of Systems and Control. Springer,
  London} .
\bibitem[{Zhang and Dong(2022)}]{ZD22}
\bibinfo{author}{Zhang, G.}, \bibinfo{author}{Dong, Z.}, \bibinfo{year}{2022}.
\newblock \bibinfo{title}{Linear quantum systems: a tutorial}.
\newblock \bibinfo{journal}{Annual Reviews in Control} \bibinfo{volume}{54},
  \bibinfo{pages}{274--294}.
\bibitem[{Zhang and James(2010)}]{zhang2010direct}
\bibinfo{author}{Zhang, G.}, \bibinfo{author}{James, M.R.},
  \bibinfo{year}{2010}.
\newblock \bibinfo{title}{Direct and indirect couplings in coherent feedback
  control of linear quantum systems}.
\newblock \bibinfo{journal}{IEEE Transactions on Automatic Control}
  \bibinfo{volume}{56}, \bibinfo{pages}{1535--1550}.
\bibitem[{Zhang and Pan(2020)}]{zhang2020dynamics}
\bibinfo{author}{Zhang, G.}, \bibinfo{author}{Pan, Y.}, \bibinfo{year}{2020}.
\newblock \bibinfo{title}{On the dynamics of two photons interacting with a
  two-qubit coherent feedback network}.
\newblock \bibinfo{journal}{Automatica} \bibinfo{volume}{117},
  \bibinfo{pages}{108978}.
\bibitem[{Zhang et~al.(2017)Zhang, Liu, Wu, Jacobs and Nori}]{Zhangjing2017}
\bibinfo{author}{Zhang, J.}, \bibinfo{author}{Liu, Y.X.}, \bibinfo{author}{Wu,
  R.B.}, \bibinfo{author}{Jacobs, K.}, \bibinfo{author}{Nori, F.},
  \bibinfo{year}{2017}.
\newblock \bibinfo{title}{Quantum feedback: Theory, experiments, and
  applications}.
\newblock \bibinfo{journal}{Physics Reports} .
\bibitem[{Zheng and Baranger(2013)}]{PRLDuke}
\bibinfo{author}{Zheng, H.}, \bibinfo{author}{Baranger, H.U.},
  \bibinfo{year}{2013}.
\newblock \bibinfo{title}{Persistent quantum beats and long-distance
  entanglement from waveguide-mediated interactions}.
\newblock \bibinfo{journal}{Physical Review Letters} \bibinfo{volume}{110},
  \bibinfo{pages}{113601}.
\bibitem[{Zhou et~al.(2017)Zhou, Chen and Shen}]{zhou2017single}
\bibinfo{author}{Zhou, Y.}, \bibinfo{author}{Chen, Z.}, \bibinfo{author}{Shen,
  J.T.}, \bibinfo{year}{2017}.
\newblock \bibinfo{title}{Single-photon superradiant emission rate scaling for
  atoms trapped in a photonic waveguide}.
\newblock \bibinfo{journal}{Physical Review A} \bibinfo{volume}{95},
  \bibinfo{pages}{043832}.
\bibitem[{Zwiller et~al.(2001)Zwiller, Blom, Jonsson, Panev, Jeppesen, Tsegaye,
  Goobar, Pistol, Samuelson and Bj{\"o}rk}]{dotsingle}
\bibinfo{author}{Zwiller, V.}, \bibinfo{author}{Blom, H.},
  \bibinfo{author}{Jonsson, P.}, \bibinfo{author}{Panev, N.},
  \bibinfo{author}{Jeppesen, S.}, \bibinfo{author}{Tsegaye, T.},
  \bibinfo{author}{Goobar, E.}, \bibinfo{author}{Pistol, M.E.},
  \bibinfo{author}{Samuelson, L.}, \bibinfo{author}{Bj{\"o}rk, G.},
  \bibinfo{year}{2001}.
\newblock \bibinfo{title}{Single quantum dots emit single photons at a time:
  Antibunching experiments}.
\newblock \bibinfo{journal}{Applied Physics Letters} \bibinfo{volume}{78},
  \bibinfo{pages}{2476--2478}.

\end{thebibliography}

\end{document}